\documentclass[%
 preprint,
 superscriptaddress,
 amsmath,amssymb,
longbibliography,
]{revtex4-2}

\usepackage{graphicx}%
\usepackage{dcolumn}%
\usepackage{bm}%
\usepackage{physics}
\usepackage{hyperref}
\usepackage{xparse}
\usepackage{color}
\usepackage{appendix}
\usepackage{enumerate}
\usepackage{booktabs}
\usepackage{pifont}
\usepackage{amsmath}
\usepackage{array, collcell}
\usepackage{lineno}
\usepackage{tabularx}
\usepackage{{makecell}}

\NewDocumentCommand{\rnl}{O{r} O{k} O{n} O{l}}{X_{#2 #3 #4} (#1)}
\newcommand{\vs}{v_\textrm{s}}

\newcommand{\RWS}{R_\textrm{VS}}

\newcommand{\T}{\tau}

\newcommand{\change}[1]{\textcolor{black}{#1}}

\newcommand{\Pe}{P_\textrm{e}}
\newcommand{\RVS}{R_\textrm{VS}}
\newcommand{\cmark}{\ding{51}}%
\newcommand{\xmark}{\ding{55}}%
\newcommand{\Pref}{P_\textrm{ref}}
\newcommand{\Pfd}{P^\textrm{fd}}
\newcommand{\Pstrr}{P^\textrm{st}_{rr}}
\newcommand{\Psttr}{P^\textrm{st}_\textrm{tr}}
\newcommand{\Pvirt}{P^\textrm{vir}_T}
\newcommand{\Pvirk}{P^\textrm{vir}_{K_{12}}}
\newcommand{\Pid}{P^\textrm{id}}
\newcommand{\Paa}{P_\textrm{nn}^\textrm{AA}}
\newcommand{\Pnaa}{P_\textrm{nn}^\textrm{no-AA}}
\newcommand{\logten}{\log_{10}}
\newcommand{\avg}[1]{\langle #1 \rangle}
\newcommand{\Ppredf}{P_\textrm{pred}^\textrm{f}}
\newcommand{\Ppredr}{P_\textrm{pred}^\textrm{r}}
\newcolumntype{D}{>{$\displaystyle}c<{$}}
\newcommand\AddLabel{%
  \refstepcounter{equation}%
  (\theequation)%
}
\newcolumntype{L}{>{\collectcell\AddLabel}r<{\endcollectcell}}

\begin{document}

\title{Physics-enhanced neural networks for equation-of-state calculations}

\author{Timothy J. Callow}
\email{t.callow@hzdr.de}
\affiliation{Center for Advanced Systems Understanding (CASUS), D-02826 Görlitz, Germany}
\affiliation{Helmholtz-Zentrum Dresden-Rossendorf, D-01328 Dresden, Germany}

\author{Jan Nikl}
\email{j.nikl@hzdr.de}
\affiliation{Center for Advanced Systems Understanding (CASUS), D-02826 Görlitz, Germany}
\affiliation{Helmholtz-Zentrum Dresden-Rossendorf, D-01328 Dresden, Germany}

\author{Eli Kraisler}%
 \email{eli.kraisler@mail.huji.ac.il}
\affiliation{Fritz Haber Center for Molecular Dynamics and Institute of Chemistry, The Hebrew University of Jerusalem, 9091401 Jerusalem, Israel}

\author{Attila Cangi}
 \email{a.cangi@hzdr.de}
\affiliation{Center for Advanced Systems Understanding (CASUS), D-02826 Görlitz, Germany}
\affiliation{Helmholtz-Zentrum Dresden-Rossendorf, D-01328 Dresden, Germany}

\date{\today}%

\begin{abstract}

Rapid access to accurate equation-of-state (EOS) data is crucial in the warm-dense matter regime, as it is employed in various applications, such as providing input for hydrodynamic codes to model inertial confinement fusion processes. In this study, we develop neural network models for predicting the EOS based on first-principles data. The first model utilizes basic physical properties, while the second model incorporates more sophisticated physical information, using output from average-atom calculations as features. Average-atom models are often noted for providing a reasonable balance of accuracy and speed; however, our comparison of average-atom models and higher-fidelity calculations shows that more accurate models are required in the warm-dense matter regime. Both the neural network models we propose, particularly the physics-enhanced one, demonstrate significant potential as accurate and efficient methods for computing EOS data in warm-dense matter. 

\end{abstract}

\maketitle

\tableofcontents

\section{Introduction}

The warm-dense matter (WDM) regime has emerged as a topic of great interest in recent years \cite{Koenig_2005,DOE09,BMPC14,falk_2018,Bonitz_WDM_review,Dornheim_WDM_review} for several reasons. Firstly, a plethora of interesting scientific phenomena occur under WDM conditions: of particular importance, given the present requirement for clean and abundant energy, is inertial confinement fusion (ICF) \cite{Kritcher_ICF_11,Lindl_ICF_04,Betti2016ICF,Craxton2015ICF}. WDM conditions also present themselves in a variety of astrophysical systems, such as stars in various stages of their life cycle \cite{brown_dwarfs,white_dwarfs,neutron_stars}, planetary cores \cite{planetary_cores_1,earth_core_iron}, and more besides \cite{Remington_HED}. Secondly, it has become possible to probe matter exposed to WDM conditions in the laboratory \cite{falk_2018,Expt_EOS_database,Redmer_XRTS,Vinko_nickel_2020,Aluminium_IPD_expt_2012,Kraus_WDM_Carbon}, due to the increasing capabilities of large experimental machines such as the National Ignition Facility \cite{NIF}, the Linac Coherent Light Source \cite{LINAC}, SACLA \cite{SACLA}, and the European XFEL \cite{XFEL}. Thirdly, the WDM regime is notoriously difficult to model, in part due to the unique position it occupies between the condensed-matter physics and plasma physics domains \cite{Dornheim_WDM_review, Bonitz_WDM_review, Callow_AA_22}.

Equation-of-state (EOS) data --- the pressure and energy of a material as a function of its temperature and mass density --- is of particular importance in WDM. For example, hydrodynamic codes that are used to guide ICF experiments rely on accurate EOS data, typically given in tabular form and interpolated to the conditions of interest, to close the conservation equations \cite{EOS_review}. Many theoretical techniques exists to compute EOS data. These include ``first-principles'' methods, namely density-functional theory molecular-dynamics (DFT-MD) \cite{Militzer_EOS_database, Hu_Be_EOS} and path integral Monte Carlo \change{(PIMC)} \cite{Militzer_EOS_database,Deuterium_PIMC}, as well as reduced models such as average-atom (AA) models \cite{Liberman_1979,starrett_aa_pressure,Pressure_warm_hot}, and extensions thereof \cite{Starrett_PMD,VAAQP}. A review of the application of these methods to EOS calculations can be found in Ref.~\cite{EOS_review}. Whilst these models differ in both accuracy and computational expense, they share the common principle of having only some fundamental physical properties, i.e. the temperature, density, and material composition, as inputs to compute the EOS. This is in contrast to alternative approaches such as the widely-used SESAME database \cite{SESAME_database}, which extrapolates EOS data from cold curves with a variety of different models.

Assessing the accuracy of EOS models is not straightforward. One of the principle difficulties in bench-marking models to experimental data is that temperature cannot be directly measured under WDM conditions \cite{Expt_EOS_database,Kraus_XRTS,Dornheim_temperature}. Historically, the temperature has been inferred from other quantities which can be measured, such as the internal energy \cite{Expt_EOS_database}, and is thus dependent on the model used for the inference; recent developments in this area hold future promise to resolve this issue \cite{moldabekov2022_thermal,Dornheim_temperature, dornheim_imaginary-time}. Therefore, whilst experimental EOS data exists, it is in short supply and not always suited for comparison with theoretical models \cite{Expt_EOS_database}.

Consequently, results from the first-principles methods, DFT-MD and PIMC, are typically taken as the highest quality benchmarks for EOS data in the WDM regime. These methods are denoted ``first-principles'' because, in theory, they exactly describe the many-body electronic structure problem. DFT-MD actually incorporates two different theories, Kohn--Sham DFT-MD (KS-DFT-MD) \cite{ks65}, which is often synonymous with DFT-MD due to its more widespread use, and orbital-free DFT-MD (OF-DFT-MD) \cite{Carter_OFDFT_intro}. Whilst both flavours are in principle exact, they rely on approximations. In KS-DFT, we have to approximate the exchange-correlation (xc) energy functional, for which there exists a wide spectrum of choices \cite{Jacobs_ladder,Cohen_Yang_challenges,XC_review}, with efforts ongoing to assess the impact of this choice in the WDM regime \cite{Karasiev_XC,Kushal_Hydrogen,Aurora_XC_1,Aurora_XC_2,Zhandos_XC_1,Zhandos_XC_2,Callow_AA_22, moldabekov2023_assessing, moldabekov2023_non-empirical}. Meanwhile, OF-DFT relies on approximations for both the kinetic energy density-functional in addition to the xc-energy \cite{Trickey_ODDFT_challenges,Trickey_OFDFT_GGA,Levy_OFDFT,Finzel_2019,Ludena_Pauli}. Likewise, PIMC calculations of real materials have historically relied on the fixed-node approximation \cite{Ceperley1991} to circumvent the fermion sign problem \cite{Troyer_fermion_sign} (although PIMC with no such approximation has recently been applied to study Hydrogen under WDM conditions \cite{Dornheim_Hydrogen_22}). Of course, both methods also rely on other fundamental assumptions (such as the Born--Oppenheimer approximation), besides various numerical approximations. 

\change{In recent years, (at least) two substantial EOS databases constructed from first-principles data have been published. The database of Militzer and co-workers \cite{Militzer_EOS_database} is composed of KS-DFT (at low temperature) and PIMC (at high temperatures) data. It includes data from a range of elements and mixtures, and is constructed from a series of historical papers. The database of Ding and Hu \cite{Hu_Be_EOS} is composed of KS-DFT (at low temperatures) and OF-DFT (at high temperatures) data. In both databases, it was established that the different methods used yielded results in close agreement in the overlap region.}

Despite the approximations used, DFT-MD and PIMC are trusted methods for computation of EOS data and thus comparisons with these methods are typically made when assessing the accuracy of simpler models, such as AA models. 
However, we have typically observed that these comparisons generally involve a relatively small set of benchmark data points (say in the region of tens of data points), and the analysis is often a visual comparison of the pressures on a log-log plot. \change{Historically, this was} most likely due to the lack of large datasets with which to make comparisons. \change{However, the EOS databases just discussed \cite{Militzer_EOS_database,Hu_Be_EOS} allow for a more comprehensive analysis of the performance of AA models.} 

A second consequence of the existence of larger databases is that they offer greater possibilities for the application of machine learning techniques such as neural networks. The use of neural networks in scientific fields such as materials science \cite{NN_physics_1,Lenz_ML_review}, quantum physics and chemistry \cite{Burke_ML_1,Burke_ML_2,DeepMind_DFT,DeepMind_schrodinger,Troyer_ML}, high-energy-density physics \cite{hatfield2021_thedata-driven}, and even in warm dense matter \cite{Dornheim_ESA,mala,fiedler2022_training-free}, is becoming increasingly popular. There are many reasons behind this, but relevant to this paper (besides the aforementioned growth in the size and number of databases) is the fact that they are excellent function approximators, with the ability to learn complex non-linear relationships between benchmark data and input features \cite{Goodfellow-et-al-2016,nns_functions}. 

Neural network models can be trained on first-principles data and then applied to arbitrary densities and temperatures within the range of the training data. Of course, interpolating EOS data is a well-established practise. However, machine learning models offer an intriguing path to interpolate EOS data with only minimal physical constraints from the user, and, for example, modern practises in machine learning can be used for more robust error estimation. Applying machine learning techniques to EOS data is still a relatively fresh topic: most of the applications to date focussed on uncertainty quantification \cite{Kraus_EOS_uncertainty,gaffney2022constraining,Lindquist_EOS_uncertainty}, although in a recent work \cite{Mentzer_EOS_neural} the authors built a surrogate EOS model using neural networks.

In this paper, our primary objective is to develop a neural network model capable of accurately interpolating the EOS by leveraging existing first-principles datasets \cite{Militzer_EOS_database,Hu_Be_EOS}. Furthermore, we explore the potential of physics-enhanced neural networks for improving model performance, by incorporating additional physical information from an AA model, which has a low computational overhead for data generation. We also show that these neural network models yield a significant improvement in performance relative to the base AA model.

The paper is structured as follows. Firstly, in Section~\ref{sec:AA_review}, we give a brief overview of the AA model which we use. Next, in Section~\ref{sec:AA_P}, we present various methods for computing the pressure in AA models, which will all be compared in the results. After that, in Section~\ref{sec:nn_method}, we explain the method that we use to train and evaluate our neural networks. Finally, in Section~\ref{sec:results}, we compare results from the AA and neural network models against the two first-principles datasets \cite{Militzer_EOS_database,Hu_Be_EOS}. We note that this section is split into separate parts: \change{in the first part, we evaluate the performance of the networks on the FPEOS database of Ref.~\cite{Militzer_EOS_database}, which was the database used to train the networks \footnote{Of course, the performance of the network is evaluated on data that was not seen during training; the point is that the training and test data is drawn from the same database}. In the second part, we evaluate the performance of the networks on the database of Ref.~\cite{Hu_Be_EOS}, which was not used in the training of the neural networks. This second part therefore tests the robustness of the networks with respect to different sources of data, and also tests their ability to interpolate to unseen species.}

\section{Average-atom models: theoretical background}\label{sec:AA_review}

Average-atom (AA) models have a long history in plasma physics \cite{AA_Wigner_Seitz,AA_Feynmann,Liberman_1979,AA_Dharma_wardana, Rozsnyai,Callow_AA_22}, especially in terms of calculation EOS data \cite{AA_EOS_1,starrett_aa_pressure,Purgatorio,Pressure_warm_hot,VAAQP}. There exists a broad range of AA models which use different assumptions and approximations. However, AA models are united by a common concept, namely the reduction of a many-body fully-interacting system of electrons and nuclei to an effective single atom model. The effect of inter-atomic interactions is coarsely accounted for via the boundary conditions used to solve the Schr\"odinger equation for the single atom. A derivation from first-principles of an AA model similar to the one used in this work can be found in Ref.~\cite{Callow_AA_22}.

We use the AA model proposed by Massacrier \emph{et al} \cite{Massacrier_bands_2021}. In this model, we solve the \change{Kohn--Sham equations in a spherically symmetric potential},
\begin{equation} \label{eq:ks}
\left[\frac{\textrm{d}^2}{\textrm{d}r^2} + \frac{2}{r}\frac{\textrm{d}}{\textrm{d}r} - \frac{l(l+1)}{r^2} \right] X_{\epsilon nl}(r) + 2 \left[\epsilon_{nl} - v_\textrm{s}[n](r) \right] X_{\epsilon nl}(r) = 0\,.
\end{equation}
\change{In the above, $X_{\epsilon nl}$ are the radial KS orbitals, and $\epsilon_{nl}$ their associated energies. The $n,l$ indices in the subscripts denote the principle and azimuthal quantum numbers. Because the average-atom represents an extended system, rather than an isolated atom, there is an energy band (for non-core states) for each pair of $n,l$ values. Within this energy band, there is a continuous set of KS orbitals with energy $\epsilon$, which is why the KS orbitals $ X_{\epsilon nl}$ have an additional subscript $\epsilon$.
}

In the KS equation \eqref{eq:ks}, $v_\textrm{s}[n](r)$ is the KS potential, given by
\begin{equation}
 v_{\textrm{s}}[n](r) = -\frac{Z}{r} + 4\pi \int_0^{\RWS} \textrm{d}{x} \frac{n(x)x^2}{\max{(r,x)}} + \frac{\delta F_\textrm{xc} [n]}{\delta n(r)}\,,
\end{equation}
The three terms in the potential are
respectively the electron-nuclear attraction, the classical Hartree
repulsion, and the exchange-correlation (xc) potential, which is equal
to the functional derivative of the xc free energy $F_\textrm{xc}[n]$. In this work, we use exclusively the ground-state (no temperature dependence) local density approximation (LDA) for $F_\textrm{xc}[n]$ \cite{ks65}, using the parameterization of Perdew and Wang \cite{PW92}. As ever, due to the dependence of the KS potential on the density \(n(r)\), the KS equations
must be solved iteratively until self-consistency is reached.

The KS equations \eqref{eq:ks} are first solved under the following two boundary conditions,
\begin{align}
0&=X_{\epsilon nl}(\RWS)\,, \label{eq:bc_dir}\\
0&=\frac{\textrm{d}X_{\epsilon nl}(r)}{\textrm{d}r}\Bigg|_{r=\RWS}\, \label{eq:bc_neu},
\end{align}
\change{where $\RVS$ is the atomic radius (or the Voronoi sphere radius), defined in terms of the ion number density $n_i$ as
\begin{equation}
    \RWS = \left(\frac{3}{4\pi n_i}\right)^{1/3}\,.
\end{equation}
Solving under the above boundary conditions} yields a set of eigen-functions $X^+_{nl}(r)$ and $X^-_{nl}(r)$ with associated eigen-energies $\epsilon_{nl}^+$ and $\epsilon_{nl}^-$. These energies define the upper and lower limits of an energy band, and the KS equations \eqref{eq:ks} are then solved for energies inside the band limits to yield $X_{\epsilon nl}(r)$, $\epsilon^-_{nl}\leq\epsilon\leq\epsilon^+_{nl}$. After discretizing the energy bands, the density $n(r)$ is equal to
\begin{gather} \label{eq:dens_masac}
n(r) = 2\sum_{k}^{N_k} w_k \sum_{nl}(2l+1) f_{knl}(\epsilon_{knl},\mu,\T) |X_{knl}(r)|^2\,,\\
w_k = \frac{8}{\pi(N_k-1)^2}\sqrt{k(N_k-1-k)}\,, \label{eq:weight_masac}
\end{gather}
where $N_k$ is the number of points used in the discretization of each energy band. The derivation of Eqs.~\eqref{eq:dens_masac} and \eqref{eq:weight_masac} can be found in Ref.~\cite{Callow_MIS}. \change{We note that we have changed the notation of the radial KS orbitals from $X_{\epsilon nl} (r)$ to $\rnl$, to signify the fact that they are represented on a discrete, rather than continuous, energy grid.}

The occupation numbers $f_{knl}$ are given in the usual way according to the Fermi--Dirac distribution,
\begin{equation}
f_{knl}(\epsilon_{knl},\mu,T) = \frac{1}{1+e^{(\epsilon_{knl}-\mu)/\T}}\,.
\end{equation}
The chemical potential \(\mu\) is determined by fixing the electron
number \(N_\textrm{e}=4\pi\int_0^{\RWS} \textrm{d}r r^2 n(r)\) to be equal to a
pre-determined value (in this paper, \(N_\textrm{e}=Z\) in all cases). \change{This is an iterative procedure because the electron density, and hence electron number, depends on the occupations $f_{knl}$, which themselves depend on the chemical potential.}

All AA calculations were performed using the open-source code atoMEC \cite{atoMEC,SciPy_atoMEC}. Ref.~\cite{SciPy_atoMEC} gives a broad overview of the code, along with a more detailed description of how to set up and run SCF calculations. Since that publication, the stress-tensor and virial methods for the pressure (see next section) have been added to atoMEC, and can be found in v1.2.0 onwards. We note that the following libraries are used extensively by atoMEC: NumPy \cite{numpy}, SciPy \cite{scipy}, LIBXC \cite{libxc_2018}, mendeleev \cite{mendeleev2014}, and joblib \cite{joblib}. 

\section{Pressure in average-atom models}\label{sec:AA_P}

The total pressure in AA models is equal to the sum of the ionic and electronic components. Typically, the ionic pressure is just given by the ideal gas pressure,
\begin{equation}\label{eq:P_ion}
    P_\textrm{ion} = \frac{n_{M}RT}{V}\,,
\end{equation}
where $n_{M}$ is the number of moles, $R$ the ideal gas constant, and $V$ the volume of the gas. Clearly, this assumption is more accurate for lower material densities and higher temperatures.

There are a number of methods to compute electronic pressure in AA models. In this paper, we consider four different methods, presented below. We note from the start that a number of papers have explored and derived links between these different formulae \cite{More_QSM,stress_tensor_more,Pressure_warm_hot,Pain_virial}, and that the stress-tensor method and the virial method should, in principle, be equivalent to the functional derivative method, which is the established thermodynamic expression for the pressure \footnote{The ideal approximation, as implied by the name, is a known approximation, and cannot be derived from the functional derivative of the free energy}. However, AA models rely on a range of different approximations, and thus an established relationship for one model is unlikely to hold for a different one. Moreover, the aim of this paper is to take a practical approach to EOS calculations, i.e. one that is informed by comparisons to first-principles datasets. Therefore, in this work we present the methods as we use them, but we do not attempt to draw theoretical links between them.

\subsection{Functional derivative of the free energy} \label{sec:P_fd}

The first method involves taking the functional derivative of the free energy with respect to the volume,
\begin{equation}\label{eq:P_fd}
   \Pfd = -\pdv{F}{V}\Bigg|_T\,,
\end{equation}
which is a well-established way to compute the pressure in thermodynamics. However, this method is used less frequently in AA calculations, because it is only thermodynamically consistent if no distinction is made between bound and unbound electrons, i.e. the same equations are solved for all orbitals, regardless of their energy. This is true in the AA model we use, but it is not the case for AA models in general. For example, in Ref.~\cite{Callow_AA_22}, in which the bound and unbound orbitals are treated differently, unusual results were observed for the pressure with this method.

In DFT, the free energy is given by
\begin{equation}
    F[n] = E[n] - T S[n],
\end{equation}
where $S[n]$ is the entropy and $E[n]$ is the internal energy functional. In KS-DFT (and thus our AA model), these terms are given by \footnote{We note that, in Eq.~(12), we have written the xc-free energy $F_\textrm{xc}[n]$ as part of the internal energy. However, in principle, it also contains an entropic contribution, because the entropy $S[n]$ is approximated by the non-interacting entropy functional in KS-DFT. We direct readers to Ref.~\cite{Callow_AA_22} for a more detailed discussion of the xc energy term in ground-state and finite-temperature KS-DFT.}
\begin{align} \label{eq:E_internal}
    E[n] &= T_\textrm{s}[n] + E_\textrm{en}[n] + E_\textrm{Ha}[n] + F_\textrm{xc}[n]\\
    S[n] &= -\sum_{k}w_k\sum_{l,n} (2l+1) \big[ f_{knl}\log(f_{knl}) + (1-f_{knl}) (\log(1-f_{knl}) \big]\,.
\end{align}
In the above, $T_\textrm{s}[n]$ denotes the KS kinetic energy, $E_\textrm{en}[n]$ the electron-nuclear attraction energy, $E_\textrm{Ha}[n]$ the Hartree energy and $F_\textrm{xc}[n]$ the xc free energy. These terms are given by
\begin{align} \label{eq:T_KS}
      T_\textrm{s}[n] &= - 2\pi  \sum_k w_k \sum_{l,n} (2l+1) f_{knl} \int_0^{\RVS} \dd{r} r^2 \rnl \dv[2]{\rnl}{r}\,,\\
        E_\textrm{en}[n] &= -4 \pi Z \int_0^{\RVS} \dd{r} r n(r)\,, \\
     E_\textrm{Ha}[n]&= \frac{1}{2}(4\pi)^2 \int_0^{\RVS} \dd{r} r^2 n(r)\int_0^{\RVS} \dd{x} \frac{n(x)x^2}{\max(r,x)}\,,\\
     F_\textrm{xc}[n] &= 4\pi  \int_0^{\RVS} \dd{r} r^2 f_\textrm{xc}[n](r)n(r)\,,   
\end{align}
\change{where $f_\textrm{xc}[n](r)$ is the xc energy density, which for LDA is the xc energy per electron of a uniform electron gas of density $n$.}

Besides the issue with thermodynamic consistency, a more minor issue with this method is that the functional derivative in Eq.~\eqref{eq:P_fd} is evaluated using finite differences, which requires two separate SCF calculations at different volumes.

\subsection{Stress-tensor}\label{sec:P_st}

We next consider the stress-tensor method for calculating the electronic pressure. This method was applied earlier using only the radial component of the stress-tensor \cite{stress_tensor_more,stress_tensor_semi_rel,stress_tensor_rel}, but more recently the full expression for the stress-tensor was used in an AA model \cite{Pressure_warm_hot}, where the AA results were seen to be in good agreement with DFT-MD calculations and experimental data. In a follow-up paper \cite{Carbon_ionization}, excellent agreement was found between an AA model and DFT-MD simulations using this method, for compressed Carbon at temperature 100 eV.

In spherical co-ordinates, the diagonal components of the stress-tensor matrix are given by
\begin{align}
    T_{rr} = \Pstrr &= \frac{1}{2} \sum_k w_k \sum_{n,l} 2 (2l+1) f_{knl} \left[ \left(\dv{\rnl}{r} \right)^2  - \rnl \dv[2]{\rnl}{r}\right]_{r=\RWS} \\
    \nonumber
    &= \frac{1}{2} \sum_k w_k \sum_{n,l} 2 (2l+1) f_{knl} \Bigg[ \left(\dv{\rnl}{r} \right)^2  + \frac{2}{r}\rnl\dv{\rnl}{r} \\
    \label{eq:P_st_rr}
    & \hspace{12em}+ \left( 2[\epsilon_{knl} - \vs(r)] - \frac{l(l+1)}{r^2} \right) X_{knl}^2(r) \Bigg]_{r=\RWS}\\
    T_{\theta\theta} = T_{\phi\phi} &= \frac{1}{2} \sum_k w_k \sum_{n,l} 2 (2l+1) f_{knl} \left[ \frac{l^2+l+1}{r^2} X_{knl}^2(r) -\frac{1}{2r^3}\dv{[r\rnl]^2}{r} \right]_{r=\RWS} \\
    &=\frac{1}{2} \sum_k w_k \sum_{n,l} 2 (2l+1) f_{knl} \left[ \frac{l(l+1)}{r^2} X_{knl}^2(r) -\frac{1}{r}\rnl\dv{\rnl}{r} \right]_{r=\RWS}\,.
\end{align}

In Ref.~\cite{stress_tensor_more} and Ref.~\cite{Pain_virial}, the radial component $T_{rr}$ is used to define the total electronic pressure, i.e. $\Pstrr= T_{rr}$. On the other hand, in Ref.~\cite{Pressure_warm_hot}, the (average of) the trace is taken, which leads to the expression
\begin{align}
    \Psttr &= \frac{1}{3}(T_{rr} + T_{\theta\theta} + T_{\phi\phi})\\
    \label{eq:P_st}
    &= \frac{1}{6}\sum_k w_k \sum_{n,l} 2(2l+1) f_{knl} \left\{ \left(\dv{\rnl}{r}\right)^2 + \left[2\left(\epsilon_{knl} - \vs(r) \right) +\frac{l(l+1)}{r^2}\right] X_{knl}^2 (r)\right\}\Bigg|_{r=\RWS}\,.
\end{align}
Eq.~\eqref{eq:P_st} is essentially the same expression as that presented in Eqs.~(17) and (18) of Ref.~\cite{Pressure_warm_hot}, but there are a few small differences. Firstly, we use the band-structure model, which introduces the weightings $w_k$, and we make no distinction between bound and free electrons, hence there is just a single term for the pressure. Additionally, we don't enforce a boundary condition on the potential, i.e. $\vs(\RWS)\neq0$ in our model; this introduces the extra potential term in Eqs.~\eqref{eq:P_st_rr} and ~\eqref{eq:P_st}. Finally, the spatial wave-functions are also defined slightly differently: \change{the conversion between $P_{nl}(r)$ in Ref.~\cite{Pressure_warm_hot} and $\rnl$ in our work is given by (noting that the orbitals $\rnl$ contain an extra index $k$ because of the band-structure model used),
\begin{equation}
    P_{nl}(r) \rightarrow \sqrt{4\pi} r \rnl\,.
\end{equation}}

However, although the formula \eqref{eq:P_st} is consistent with Ref.~\cite{Pressure_warm_hot}, the individual components $P_{rr}$ and $P_{\theta\theta}$ are, in fact, different. These differences cancel out when the trace is taken. Of course, this does not affect the results in Ref.~\cite{Pressure_warm_hot} because the individual components are not used in that paper, but we mention it here to avoid confusion for the reader. When we present results in Sec.~\ref{sec:results}, we consider both the radial component, $\Pstrr$ \eqref{eq:P_st_rr}, and the trace, $\Psttr$.

\subsection{Virial theorem}\label{Sec:P_vir}

Usually the virial theorem relates the pressure to the kinetic energy $T$ and potential energy $U$ as follows \cite{virial_slater,virial_mclellan,virial_nielsen}:
\begin{equation}\label{eq:P_vir}
    \Pvirt = \frac{2T + U}{3V}\,,
\end{equation}
where the potential energy $U$ denotes the sum of all the interaction energies in the system. In DFT, the above expression becomes \cite{PBE_virial}
\begin{equation}\label{eq:P_vir_KS}
    \Pvirt = \frac{2 T_\textrm{s} + E_\textrm{en}+E_\textrm{ha} + W_\textrm{xc}}{3V}\,.
\end{equation}
The form of $W_\textrm{xc}$ depends on the type of xc-functional used. For the LDA xc-functional used in this paper, it is given by \cite{PBE_virial}
\change{
\begin{equation}
    W_\textrm{xc}^\textrm{LDA} = -3 \left[ F_\textrm{xc}^\textrm{LDA} + 4\pi\int_0^{\RWS} r^2 n(r) v_\textrm{xc}^\textrm{LDA}(r) \right]\,.
\end{equation}}

However, it is noted, e.g. in Refs.~\cite{stress_tensor_more} and \cite{Pain_virial}, that the expressions \eqref{eq:P_vir} and \eqref{eq:P_vir_KS} are only valid in the case of an infinite system ($\RWS\to\infty$), or if \change{\emph{all} the wave-functions obey one of the boundary conditions \eqref{eq:bc_dir} or \eqref{eq:bc_neu}}.
Neither of these assumptions is true for the AA model we use \change{because the boundary conditions \eqref{eq:bc_dir} and \eqref{eq:bc_neu} only determine the band-structure limits, and we solve for all wave-functions between these limits}. Instead, Refs.~\onlinecite{stress_tensor_more} and \onlinecite{Pain_virial} propose the following form for the virial pressure,
\begin{equation}\label{eq:P_vir_alt}
    \Pvirk = \frac{K_1 + K_2 + U}{3V}\,.
\end{equation}
In Eq.~\eqref{eq:P_vir_alt}, the first term $K_1$ is equal to the kinetic energy defined in Eq.~\eqref{eq:T_KS}, $K_1=T_\textrm{s}$. The second term $K_2$ is given by
\begin{equation}
    K_2= 2\pi  \sum_k w_k \sum_{l,n} (2l+1) f_{knl} \int_0^{\RVS} \dd{r} r^2 \left\{\left| \dv{\rnl}{r} \right|^2 + \frac{l(l+1)}{2r^2}X_{knl}^2(r) \right\}\,.
\end{equation}
It is straightforward to see that $K_1=K_2=T_\textrm{s}$ for the case when either of the conditions \eqref{eq:bc_dir} or \eqref{eq:bc_neu} holds. In fact, $K_2$ is a well-known alternative expression for the kinetic energy in KS-DFT \cite{Ayers_KED,Cohen_1979}, and is used, for example, to calculate the electron localization function \cite{Savin_ELF_1996}.

In Refs.~\onlinecite{Pain_virial} and \onlinecite{stress_tensor_more}, it was shown that the expression \eqref{eq:P_vir_alt} for the virial pressure, and the radial component of the stress-tensor \eqref{eq:P_st_rr}, are in principle equivalent. However, as discussed, AA models are based on various different assumptions which may mean that an established relationship in one model does not hold for another. For example, in the proof presented in Ref.~\onlinecite{stress_tensor_more}, there is no explicit exchange-correlation term included in the virial formula.

In this work, we calculate the pressure with both forms of the virial expression $\Pvirt$ \eqref{eq:P_vir} and $\Pvirk$ \eqref{eq:P_vir_alt} and benchmark the results against first-principles simulations.

\subsection{Ideal approximation}\label{sec:P_id}

The final method we use to compute the pressure in the AA model is based on the assumption that the electron density at the sphere boundary is completely free \cite{Johnson_presssure}. We therefore call this approach the ideal approximation, and the electron pressure is calculated as \cite{Callow_AA_22}
\change{\begin{equation}\label{eq:P_id}
    \Pe^\textrm{id} = \frac{2^{3/2}}{3\pi^2} \int_{v_\textrm{s}(\RWS)}^\infty \dd{\epsilon}
    \frac{\epsilon^{3/2}}{1+e^{(\epsilon-\mu)/\T}}\,.
\end{equation}
The lower bound of the integral is the value of the KS potential at the atomic radius, $v_\textrm{s}(\RWS)$, instead of zero. This is because the KS potential is not zero at the boundary, and electrons are considered ``free'' if their energy exceeds the KS potential at the boundary.}

In Ref.~\cite{Pressure_warm_hot}, it is shown that the above expression for the electron pressure is consistent with both the trace of the stress-tensor \eqref{eq:P_st} and the radial component \eqref{eq:P_st_rr}, under the assumption of free electron density. Of course, it is unclear how well the assumption of free electron density at the sphere boundary holds, particularly considering the known difficulties of defining whether electrons are free or bound in the WDM regime \cite{Callow_MIS}. However, since the AA model is already built on numerous approximations, the accuracy of the above expression will be assessed in the results section.

\section{Neural network methodology}\label{sec:nn_method}

\begin{table}[]
    \centering
    \renewcommand{\arraystretch}{1.2}    
    \begin{tabular}{DcDL}
    \toprule
         \text{Name (formal)} & Name (informal) & \text{Definition} & \multicolumn{1}{l}{} \\ \midrule
        \epsilon_1(y_1, y_2) & (M)APE$^{1}$ &  100 \times \Bigg|\frac{y_1 - y_2}{y_1}\Bigg| & \label{eq:MAPE} \\
        \epsilon_2(y_1, y_2) & (M)ALE$^{2}$ & |\logten y_{1} - \logten y_{2}| & \label{eq:MALE} \\
        \epsilon_3(y_1, y_2) & ad(M)ALE$^{3}$ & |\logten y_1 (\logten y_{1} - \logten y_{2})| & \label{eq:adMALE} \\
        f_{20} & $f_{20}$ & 100 \times \frac{\mathcal{N}(\epsilon_1 \leq 20)}{\mathcal{N}(\epsilon_1)} & \label{eq:f20}\\
        f_{5} & $f_{5}$ & 100 \times \frac{\mathcal{N}(\epsilon_1 \leq 5)}{\mathcal{N}(\epsilon_1)} & \label{eq:f5} \\  
    \bottomrule
    \end{tabular}
    \caption{The error metrics that are used in this work. We note that the first three metrics, $\epsilon_i(y_1, y_2)$, $i=1..3$, denote an error measurement between two data points $y_1$ and $y_2$. If the mean (M) over a set of data points is computed, this gives an average error. The last two error metrics, $f_{20}$ and $f_5$, which represent the fraction of points with under $20\%$ and $5\%$ errors respectively, are always aggregate measures. \\
    $^1$ \footnotesize{(Mean) absolute percentage error} \\
    $^2$ \footnotesize{(Mean) absolute log error} \\
    $^3$ \footnotesize{Adjusted (mean) absolute log error}
    }
    \label{tab:error_metrics}
\end{table}

There are various flavours of neural network, such as convolutional neural networks \cite{CNNs_1, CNNs_2, CNNs_crystals}, recurrent neural networks \cite{RNNs_1}, and generative adversarial networks \cite{GANs}. In our work, we use a standard multi-layer perceptron (MLP) feed-forward neural network \cite{Rumelhart1986}. As mentioned in the introduction, we have trained two completely separate neural networks. The first network is trained using only fundamental physical quantities --- i.e., the temperature, material density, and atomic number $Z$ --- together with the ideal gas pressure, Eq.~\eqref{eq:P_ion}, as input features. The second network is trained using various outputs of an AA calculation, chosen based on physical intuition, in addition to the aforementioned physical quantities, as input features. Aside from this difference, the networks were trained in an identical manner.

We follow a nested cross-validation workflow to train and evaluate our networks \cite{hastie_stats}. In this procedure, there is an inner cross-validation loop during which feature selection and hyper-parameter optimization are performed; and an outer loop, in which the generalization error (the performance of the network on unseen data) is evaluated. Nested cross-validation is a computationally expensive procedure, however it is recommended for relatively small datasets (such as the one used in this paper), because it avoids overfitting to the training sets (bias) and yields a robust estimate for the generalization error.

In this paper, we work with several error evaluation metrics. These are used not only in the neural network training and evaluation, but also in evaluation of the base AA results. As shall become apparent in Sec.~\ref{sec:results}, it is difficult to compare the performance between different models using just a single error metric, which is why we consider several in this work. In Table~\ref{tab:error_metrics}, we present the error metrics used. The (mean) absolute percentage error (MAPE) \eqref{eq:MAPE} and (mean) absolute log error (MALE) \eqref{eq:MALE} are widely-used expressions, well-suited to this work because the range of output values (pressures) spans many orders of magnitude (making, for example, the mean-squared error a useless metric). The adjusted (mean) absolute log error (adMALE) \eqref{eq:adMALE} is similar to the MALE but gives greater weighting to higher pressures, which (as we shall later see) is a helpful property for network training. Finally, the $f_{20}$ \eqref{eq:f20} and $f_5$ \eqref{eq:f5} scores are useful metrics because they are not sensitive to the presence of large outliers.

Below, we outline the workflow we use to train and test our neural networks. As discussed, the same procedure is used for the networks trained with and without AA features, with the only difference being the set of features that can be used in the training. In Secs. \ref{sec:feature_scaling} and \ref{sec:nn_architecture}, we discuss in more detail the feature selection and hyper-parameter optimization procedures. Note that, below and throughout this paper, we use the notation $\Pref$ and $Y^0$ to denote the reference (ground-truth) pressure values \change{\footnote{The reason we use two different notations is because $\Pref$ is used to denote the reference pressure in general, and is used later when evaluating the raw AA results, as well as the neural network results.  $Y^0$ is only used in the context of the neural network training procedure, and is used to denote a specific reference pressure for the given stage / subset of the training workflow. Furthermore, as will be discussed in Section \ref{sec:feature_scaling},  $Y^0$ is a transformation of the original pressure via a scaling relation.}}, \change{i.e. the first-principles datasets of Ref.~\onlinecite{Militzer_EOS_database} (used for network training and evaluation) and Ref.~\onlinecite{Hu_Be_EOS} (used for network evaluation only).}

\begin{enumerate}
    \item Assign class-labels $C_i,\ i=1..10$ based on the magnitude of the reference pressure $P_\textrm{ref}$, i.e. the lowest 10\% values of $P_\textrm{ref}$ are labelled $C_1$, and so on.
    \item Using a five-fold stratified cross-validation (SCV) approach, split the dataset randomly (preserving equal ratios of the class labels in each splitting) into training and test sets, $S^\textrm{tr}_i$ and  $S^\textrm{te}_i,\ i=1..5$. Put the test sets aside.
    \item For each training set $S^\textrm{tr}_i$, randomly select a set of features $\{f_m\}$
    \item For each set of features $\{f_m\}$:
    \begin{enumerate}[i]
        \item Pick a set of hyper-parameters $\{h_k\}$.
        \item Split the training set $S^\textrm{tr}_i$ using five-fold SCV into new training and validation sets, $\tilde{S}^\textrm{tr}_{ij}$ and  $\tilde{S}^\textrm{te}_{ij},\ j=1..5$.
        \item Train a network for each training set $\tilde{S}^\textrm{tr}_{ij}$, with the chosen hyper-parameters and features. Let us denote these networks $g_{ijkm}[S]$, where $S$ is the data (or dataset) on which they are evaluated.
        \item Compute the adMALE $\avg{\epsilon_{3,ikm}}$ over $j=1..5$ for the given hyper-parameters and features:
 \begin{equation}\label{eq:adjALE_ikm}
      \avg{\epsilon_{3,ikm}} = \frac{1}{N^\textrm{tr}_{i}}\sum_{j=1}^5\sum_{n=1}^{N^\textrm{te}_{ij}} \epsilon_3(Y^0_{ijn}, g_{ijkm}[\tilde{S}^\textrm{te}_{ijn}])
 \end{equation}
        where $N^\textrm{tr}_{i}$ is the total number of samples in the training set $\tilde{S}^\textrm{tr}_{i}$, $N^\textrm{tr}_{i}=\sum_{j=1}^5 N^\textrm{te}_{ij}$, and $n$ denotes a sample within that set.
        We note that all the reference pressures $Y^0$ are positive, and because we use the rectified linear activation function in the neural networks, all the predictions are also strictly positive. Therefore there are no problems taking the logarithms.
        \item Pick a new set of $\{h_k\}$. Repeat steps ii-vi 10 times, and finally save the minimum error metric and the associated hyper-parameters:
        \begin{equation}\label{eq:M0ik}
            \avg{\bar{\epsilon}^0_{3,im}} = \min_{\{h_k\}} \avg{\epsilon_{3,ikm}}
        \end{equation}
    \end{enumerate}
    \item Repeat steps 3-4 20 times, saving the error metric each time and features used for each repetition. At the end, select the \emph{three} lowest error metrics and save their associated features:
    \begin{align}
        \avg{\bar{\epsilon}_{3,i}^0} &= \min_{\{f_m\}}  \avg{\bar{\epsilon}^0_{3,im}} \\
         \avg{\bar{\epsilon}_{3,i}^1} &= (\min_{\{f_m\}}+1) \avg{\bar{\epsilon}^0_{3,im}}  \\        
         \avg{\bar{\epsilon}_{3,i}^2} &= (\min_{\{f_m\}}+2) \avg{\bar{\epsilon}^0_{3,im}} \,, 
    \end{align}
     where $(\min+1)$ and $(\min+2)$ denote the second and third lowest errors respectively. We shall comment on this step in more detail below, because it is not a standard part of a nested cross-validation (CV) procedure.
    \item Following steps 3-5, we end up with three optimal sets of features $\{f_m^l\}$ and hyper-parameters $\{h_k^l\}$, $l=1,2,3$, for the $i$-th iteration of the outer CV loop. Using these optimal sets, train three networks with the \emph{full} training set $S_i^\textrm{tr}$. We denote these networks $g_{il}[S]$.
    \item Using the networks trained in the above step, make predictions on the corresponding \emph{test} dataset $S_i^{\textrm{te}}$. The final prediction, which we denote $\bar{g}_i[S]$, is given by a simple average of the three individual predictions,
    \begin{equation}
        \bar{g}_i[S] = \frac{1}{3} \sum_{l=0}^2 g_{il}[S].
    \end{equation}
    With these predictions, evaluate various error metrics $\avg{\bar{\epsilon}_{j,i}}$ on the corresponding \emph{test} dataset $S_i^{\textrm{te}}$. Finally, take the average over all the test datasets to compute the final error metrics.
For example, the MAPE \eqref{eq:MAPE}, after averaging over all test sets, is defined as
    \begin{equation}
        \avg{\bar{\epsilon}_{1}} = \frac{1}{N_\textrm{tot}}\sum_{i=1}^5 \sum_{n=1}^{N_i^\textrm{te}} \epsilon_1\left(Y^{0,\textrm{te}}_{in}, \bar{g}_i[S^\textrm{te}_{in}]]\right),
   \end{equation}
   where $N_\textrm{tot}=\sum_{i=1}^{5}N_i^\textrm{te}$ is the total number of samples in the full dataset.
   \item The above steps tell us the \emph{generalization} error as opposed to yielding a specific model. To train the final model, repeat steps 4-6, but instead of looping over separate training sets $S_i^\textrm{tr}$, use the \emph{full} dataset.
\end{enumerate}

We emphasize that steps 1 and 2 are carried out once and only once, before any network training or feature selection begins. This ensures there is no contamination of training and test data. 

As promised above, we now elaborate a bit on step 5, in which the best three combinations of features and hyper-parameters were chosen. Typically, only a single set of features and hyper-parameters (that with the lowest error) would be carried forward in this stage. However, in order to reduce the bias (the likelihood of over-fitting to a specific model), we take forward the three best networks and then later use them to make an average prediction. This idea is borrowed from the idea of ensemble models in machine-learning \cite{Kittler1998}, where multiple models are trained and the outputs from those models are used to build a new model; however, our case is a very simple example because we just do a basic averaging in the final step, without any optimization on the inner validations sets.

\subsection{Feature selection and scaling}\label{sec:feature_scaling}

Initially, a set of possible features was chosen for the two networks. For the AA-free network, we considered only some fundamental physical quantities: namely, the material density and temperature, the volume of the Voronoi sphere, the atomic number, and the ideal gas pressure for the ions \eqref{eq:P_ion}. For the AA network, various outputs from the AA model were considered as features: for example, the electronic pressure computed using the four different methods described in Section~\ref{sec:AA_P}, the value of the density at the Voronoi sphere boundary, the mean ionization state, and so on. These features were selected from a physically intuitive viewpoint, because there is reason to expect a correlation between the feature and the reference pressure.

An initial thinning of the features was performed by hand. Using the first training set following the outer SCV splitting, $S_1^\textrm{tr}$, we first computed the correlation of each feature and the reference pressure, using the Kendall--Tau correlation measure \cite{kendall_tau}. The results (in descending order of correlation) are shown in Table~\ref{tab:init_features}. Following this, the three features with the lowest correlations, namely the atomic number $Z$, the chemical potential $\mu$, and the free energy $F[n]$, were dropped (there were additional reasons for dropping $F[n]$ and $\mu$, \change{which are discussed in the paragraph following Fig.~\ref{fig:feature_correlation}.}

Next, the correlations between the different features were checked. The potential $v_\textrm{s}(\RWS)$ and electron density $n(\RWS)$ at the cell boundary were found to be fully anti-correlated ($C=-1$); in fact, this is expected due to the Hohenberg--Kohn theorem in DFT, which establishes a one-to-one mapping between density and potential \cite{hk64}. As a result, $v_\textrm{s}(\RWS)$ was excluded from the list of potential features. Although the correlations between the different AA electron pressures were high ($C\geq 0.9$), we kept all of them as potential features as their correlation with the reference pressure was also high.

\begin{table}[]
\begin{tabular}{cccc}
    \toprule
     Feature & AA network & AA-free network & Correlation to target \\
     \midrule
     $\Pstrr$ & \cmark & \xmark & 0.928 \\
     $\Pid$ & \cmark & \xmark & 0.927 \\
     $\Psttr$ & \cmark & \xmark & 0.925 \\
     $\Pfd$ & \cmark & \xmark & 0.919 \\
     $\Pvirt$ & \cmark & \xmark & 0.916 \\      
     $\Pvirk$ & \cmark & \xmark & 0.900 \\         
     $\dv*{n(r)}{r}|_{\RWS}$ & \cmark & \xmark & 0.844 \\
     $P_\textrm{ion}$ & \cmark$^*$ & \cmark & 0.834 \\
     $\dv*{v_\textrm{s}(r)}{r}|_{\RWS}$ & \cmark & \xmark & 0.749 \\
     ${n(\RWS)}$ & \cmark & \xmark & 0.687 \\
     ${v_\textrm{s}(\RWS)}$ & \xmark & \xmark & 0.687 \\
     $T$ & \cmark & \cmark & 0.661 \\
     $\rho_\textrm{m}/Z$ & \cmark & \cmark & 0.540 \\
     $V$ & \cmark & \cmark & 0.531 \\
     $Z^*$ & \cmark & \xmark & 0.502 \\
     $\rho_\textrm{m}$ & \cmark & \cmark & 0.490 \\
     $F[n]$ & \xmark & \xmark & 0.434  \\ 
     $\mu$ & \xmark & \xmark & 0.238 \\
     $Z$ & \xmark & \xmark & 0.116 \\
     \bottomrule
\end{tabular}
    \caption{Initial features considered for the AA and AA-free neural networks, and the Kendall-Tau correlation figures between the features and the reference pressure. We also show which features are used for both networks during the training.\\
    $^*$ \footnotesize{Summed together with the electron pressures in the network with AA features}}
    \label{tab:init_features}
\end{table}

In Table~\ref{tab:init_features}, we show which features remained and which were discarded following this manual thinning of the feature space, for the AA and AA-free networks. For the remaining features, we plotted the relationship between the reference pressure and feature, to get a better insight into their relationship. We observed that, for every feature considered, a much better correlation could be observed between the feature and reference pressure when logarithms of both were taken. A couple of examples of the features plotted against the pressure, with and without taking logarithms, can be found in Fig.~\ref{fig:feature_correlation}. \change{ We note that, in this table, $Z^*$ denotes the mean ionization state (MIS), which is defined here as
\begin{equation} \label{eq:MIS}
    Z^* = 2\sum_{k} w_k \sum_{n,l} (2l+1) f_{knl}(\mu, \T) \Theta \big(\epsilon_{knl} - v_\textrm{s}(\RWS)\big)\,,
\end{equation}
where $\Theta\big(\epsilon_{knl} - v_\textrm{s}(\RWS)\big)$ is the Heaviside step function. In other words, the MIS is defined as the number of electrons with energy above the potential energy at the sphere boundary.}

We note that the AA pressures sometimes have a negative sign, but in these limited instances the absolute value of the pressure was taken instead. All the other features considered had a consistent sign; this further motivated the dropping of the chemical potential and free energy as features, since they spanned a wide spectrum of positive and negative signs for the dataset used, which complicated the use of logarithms.

\begin{figure}
    \centering
    \includegraphics{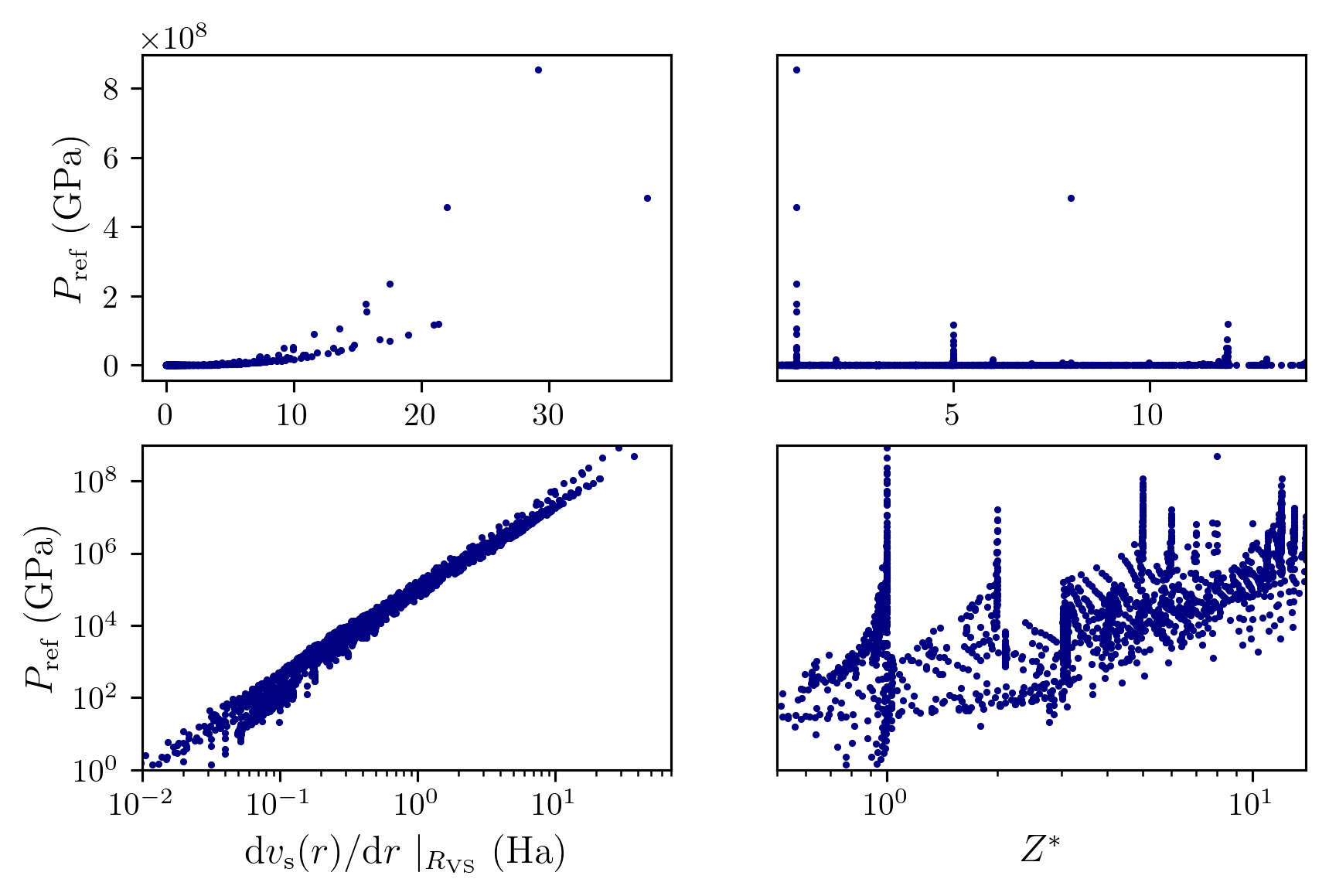}
    \caption{Features (output from AA calculation) vs reference pressure. Top row: linear scale. Bottom row: logarithmic scale. \change{Left column: derivative of KS potential at sphere boundary. Right column: MIS \eqref{eq:MIS}.} This plot indicates the potential benefits of using a logarithmic scale for both the features and the reference pressure.}
    \label{fig:feature_correlation}
\end{figure}

Furthermore, some preliminary networks were trained using the training set $S^\textrm{tr}_1$. These networks were not trained using the full workflow described in the previous sub-section. Rather, a few simple networks --- consisting of one or two hidden layers and between 10-30 neurons per layer --- were trained (also varying a few other hyper-parameters), just to get an initial idea of whether the logarithmic scaling improved the network performance. The outcome of this investigation clearly indicated that taking the logarithm of the reference pressure and all features yields significantly lower errors for all the different network architectures. Consequently, this logarithmic scaling was adopted for all subsequent training of networks.

As described in the previous sub-section, in step 3 of the workflow a random subset of features is chosen from the full feature space. In total, there were 14 available features for the AA network (after the manual reduction performed above) and just 5 for the AA-free network. For the AA network, we randomly selected between 5 and 13 features in step 3. As discussed, following hyper-parameter optimization for this set of features, the process was repeated; in total, it was repeated 20 times for the AA network. For the AA-free network, since the number of features is limited, we randomly chose between 3-5 features, and again repeated the process 20 times. Of course, there is only one subset of 5 features in this case, but repeating the hyper-parameter optimization with this same set of features is still beneficial, because the hyper-parameter search will change every time.

\begin{figure}
    \centering
    \includegraphics{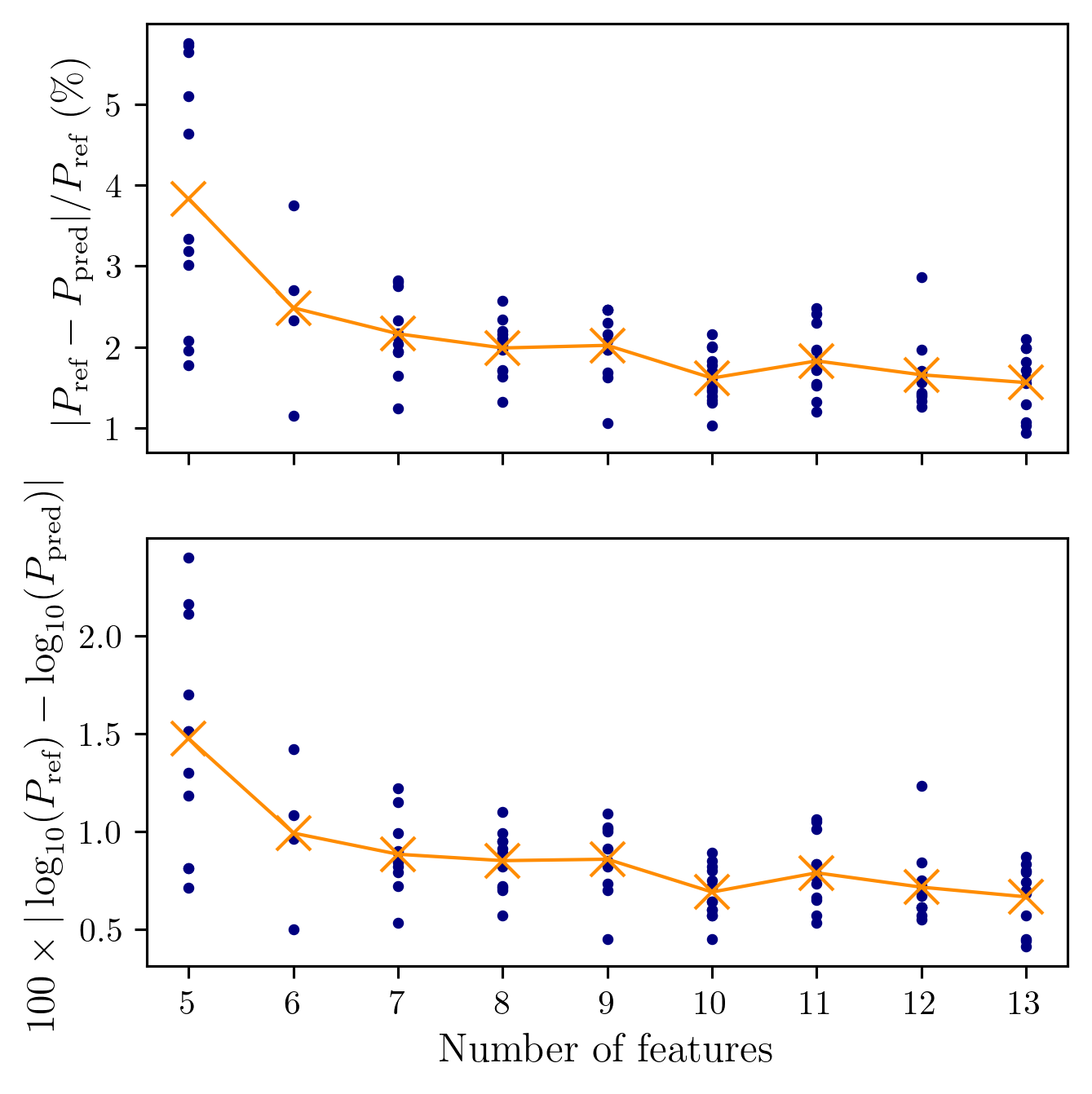}
    \caption{Mean absolute percentage error (MAPE) (top) and mean absolute log error (bottom) for the AA neural networks trained with different numbers of features. The average for each number of features is shown with the orange line.}
    \label{fig:feature_errors}
\end{figure}

In Fig.~\ref{fig:feature_errors} we plot the error metrics for the feature-sampling procedure described above, for the AA network. Note that each of the dark blue points in this graph denotes the error $\avg{\bar{\epsilon}_{3,im}^0}$ \eqref{eq:M0ik}, i.e. the error for that particular feature subset, following the inner CV search for the best hyper-parameters. This plot is an amalgamation of all the outer loop training sets $S_i^\textrm{tr}$. The dark orange line tracks the average (mean) error over the number of features selected for the sample. This information is also summarized in Table~\ref{tab:feature_errs}. We observe that, in general, more features leads to better performance; \change{however, the best performing models actual have similar errors, regardless of the number of features used.}

Following the logarithmic scaling described above, we scaled both the features and the reference pressure for each of the training sets $S_\textrm{i}^\textrm{tr}$. This was done via the standard feature scaling formula,
\begin{equation}
    \tilde{f}_{nm} = \frac{f_{nm} - u_m}{s_m},
\end{equation}
where $f_{nm}$ and $\tilde{f}_{nm}$ are the original and scaled features for the $n$-th data point, and $u_m$ and $s_m$ are respectively the mean and standard deviation of the $m$-th feature. The same approach was used for the reference pressure. Feature scaling is a standard procedure in machine learning; on the other hand, the target is not usually scaled. However, the preliminary networks we trained performed much better when the target was also scaled, and therefore we adopted this approach.

\begin{table}[]
    \centering
    \begin{tabular}{ccc}
    \toprule
         Feature number & MAPE (\%) & MALE  \\
         \midrule
         5 & 3.8 & 0.015 \\
         6 & 2.5 & 0.0099 \\
         7 & 2.2 & 0.0088 \\
         8 & 2.0 & 0.0085 \\
         9 & 2.0 & 0.0086 \\
         10 & 1.6 & 0.0069 \\
         11 & 1.8 & 0.0079 \\
         12 & 1.7 & 0.0071 \\
         13 & 1.6 & 0.0066 \\\bottomrule
    \end{tabular}
    \caption{Average mean absolute percentage error (MAPE) and mean absolute log error (MALE) for different numbers of features, over the \emph{training} sets (for the AA neural network).}
    \label{tab:feature_errs}
\end{table}

\subsection{Network architecture and hyper-parameter optimization}\label{sec:nn_architecture}

As mentioned earlier, we use a standard feed-forward neural network architecture. As discussed in the previous sub-section, we trained some initial networks using just the training set $S_1^\textrm{tr}$ to get a general idea of what features and hyper-parameters to use in our networks. This initial exploration indicated which hyper-parameters should be fixed, and of the hyper-parameters that were not fixed, what range of values to consider during the automated hyper-parameter optimization phase.

Following this initial training, we decided to use the `Adam' stochastic optimization method \cite{Adam_optimizer} and the rectified linear (ReLU) activation function, $\textrm{ReLU}(x) = \max(x,0)$. Below, we summarize the hyper-parameters that we allowed to vary during the automated search for the best hyper-parameters:
\begin{itemize}
    \item The number of hidden layers in the network, $N_\textrm{hidden}$
    \item The number of neurons in the \emph{first} hidden layer, $N_\textrm{neuron}^{(1)}$
    \item The fraction by which the number of neurons in successive layers are reduced in size, $\alpha_r$; for example, the second hidden layer has $N_
    \textrm{neuron}^{(2)} = \alpha_r N_\textrm{neuron}^{(1)}$ neurons
    \item The number of epochs over which training is performed, $N_\textrm{epoch}$
    \item The size of the mini-batch used for training, $N_\textrm{batch}$
    \item The learning rate $l_r$
\end{itemize}
All remaining hyper-parameters were fixed to their default arguments in the TensorFlow library \cite{tensorflow2015-whitepaper}.

With the hyper-parameters listed above, the Optuna library \cite{optuna_2019} was used to search for the best hyper-parameters. For each hyper-parameter, a search window was specified (based on observations from the preliminary networks that were trained); the search windows that were used are shown in Table~\ref{tab:hyperparam_search}. As mentioned, the optimal hyper-parameters were chosen as those which minimize the adMALE  \eqref{eq:adjALE_ikm} between the prediction and reference pressures for a given set of features. This error metric is related to the mean-absolute log error (MALE), however, we scale the errors with the logarithm of the reference pressure. This scaling is done because the raw AA models perform best at high pressures (resulting from high-temperatures and densities), and it would be undesirable to train a neural network that performs badly under these conditions. This scaling gives a slightly bigger weighting to higher-pressure errors.

On the other hand, the loss function used in the network training was the standard mean-absolute log error (MALE) between the reference and predicted pressures,
\begin{equation}\label{eq:Likm}
    L_{ijkm} = \frac{1}{N_{ij}^\textrm{tr}}\sum_{n=1}^{N_{ij}^\textrm{tr}}\epsilon_2(Y^0_{ijn}, g_{ijkm}[\tilde{S}_{ijn}^\textrm{tr}])\,.
\end{equation}
This was chosen (as opposed to the weighted MALE described above) because the MAE is a regularly-used loss function in neural networks, and therefore is implemented in an optimal way in neural network libraries. However, there is an added bonus: by using the MALE as the loss function and the weighted MALE as the error metric for hyper-parameter optimization and feature selection, we actually minimize (to some extent) two important errors.

We finish this section by briefly summarizing the training and evaluation procedures for our neural network models, and their architectures. Our neural network models are based on a standard feed-forward (MLP) network architecture \cite{Rumelhart1986}. They are shallow networks, with a maximum of two hidden layers used during the hyper-parameter optimization. We train two kinds of networks, one whose input features are just basic physical properties, and another which uses additional features from the output of an AA calculation. We use a nested CV approach \cite{hastie_stats}: this means hyper-parameter optimization and feature selection are performed on an inner CV loop, meanwhile errors are evaluated on an outer CV loop, which ensures no contamination of training and test data. Several error metrics, shown in Table~\ref{tab:error_metrics}, are calculated. Borrowing from the idea of ensembles in machine learning \cite{Kittler1998}, final predictions are constructed by taking the average predictions of the three best models. 

\begin{table}[]
    \centering
    \begin{tabular}{ccc}
        \toprule
         Hyper-parameter & Min value & Max value \\ \midrule
         $N_\textrm{hidden}$ & 1 & 2 \\ 
         $N_\textrm{neuron}^{(1)}$ & 20 & 80 \\
         $\alpha_r$ & 0.5 & 1.0 \\
         $N_\textrm{epoch}$ & 1000 & 2000 \\
         $N_\textrm{batch}$ & 20 & 60 \\
         $l_r$ & $10^{-4}$ & $10^{-3}$ \\ \bottomrule
    \end{tabular}
    \caption{Search windows for hyper-parameters}
    \label{tab:hyperparam_search}
\end{table}

\section{Results}\label{sec:results}

\subsection{FPEOS database of Militzer \emph{et al.}: Average-atom results}

\begin{figure}
    \centering
    \includegraphics{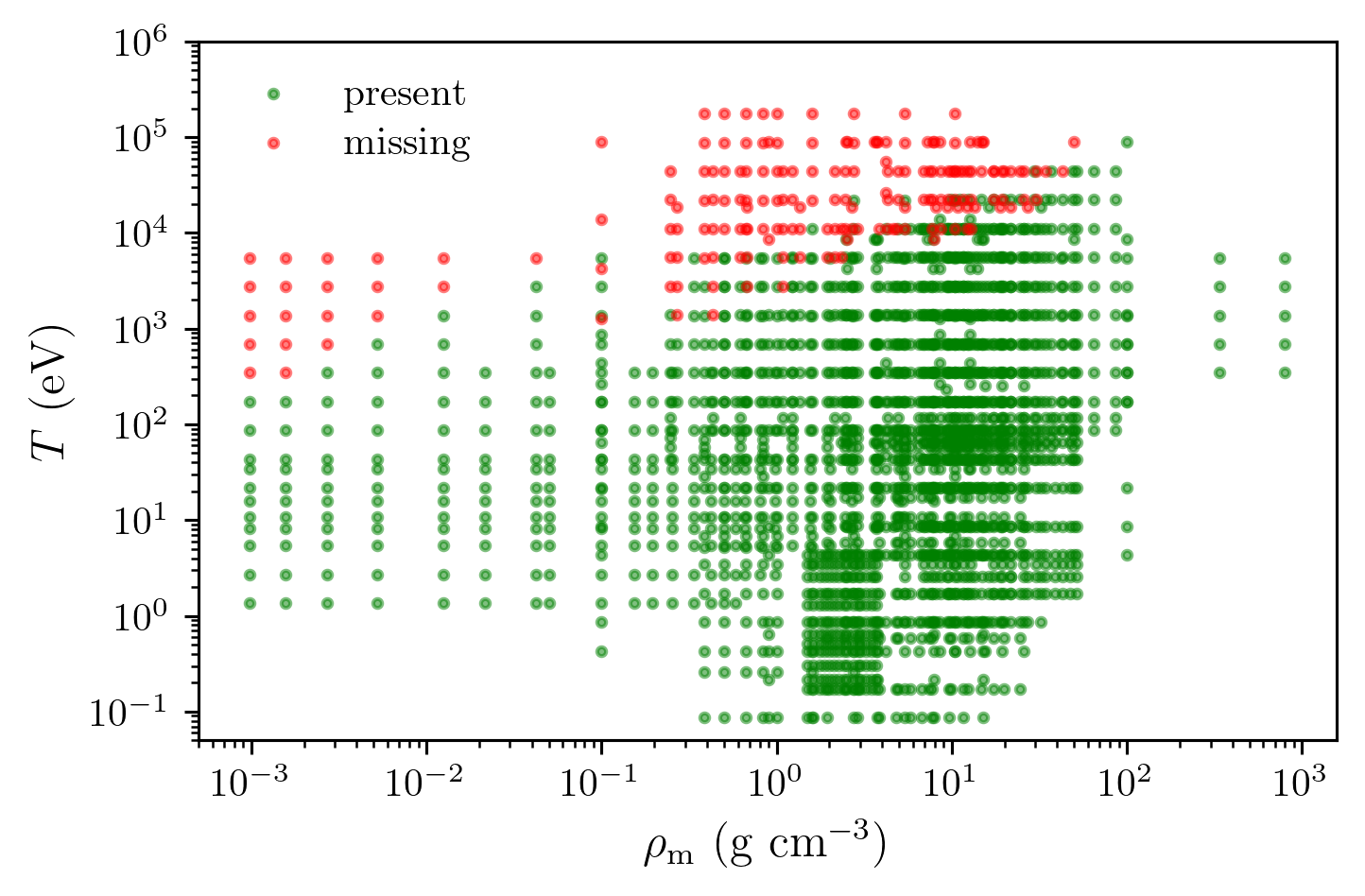}
    \caption{The data-points from the FPEOS database \cite{Militzer_EOS_database} that could be calculated with the atoMEC code (shown in green), and those that were inaccessible (in red), in terms of the density and temperature.}
    \label{fig:aa_missing}
\end{figure}

We start by presenting the results of the AA model, using the four different approximations for the pressure described in Sec.~\ref{sec:AA_P}, compared to the FPEOS dataset of Militzer \emph{et al.} \cite{Militzer_EOS_database}. From this dataset, we use only single-element data, with atomic number $1\leq Z\leq 14$. We do not use the mixtures from this dataset, because at present our AA method is not suited for mixture calculations. Furthermore, for a small subset of the conditions (high temperatures and low densities), we were unable to perform AA calculations. This is a limitation of the atoMEC code, which uses sparse matrix diagonalization to solve the KS equations \cite{SciPy_atoMEC}; under these conditions, the number of eigenvalues (KS orbitals) required is so large that the sparse matrix diagonalization breaks down. In general, however, these conditions are accessible without issues for AA codes. Nevertheless, this still leaves us with \change{2181} total data points, out of a possible 2371 had we been able to perform all calculations. \change{In Fig. \ref{fig:aa_missing}, we show the data points which we were able to compute with the AA model (in green), and those which were inaccessible (in red). As a brief aside, we note that the same set of 2181 points was used in the training of both neural networks (Section~\ref{sec:results_nn_fpeos}). In principle, all the data points from the FPEOS database \cite{Militzer_EOS_database} could be used to train the network which does not use AA features; however, we choose to use the same subset to facilitate comparison between the two networks and the raw AA results.}

\begin{figure}
    \centering
    \includegraphics{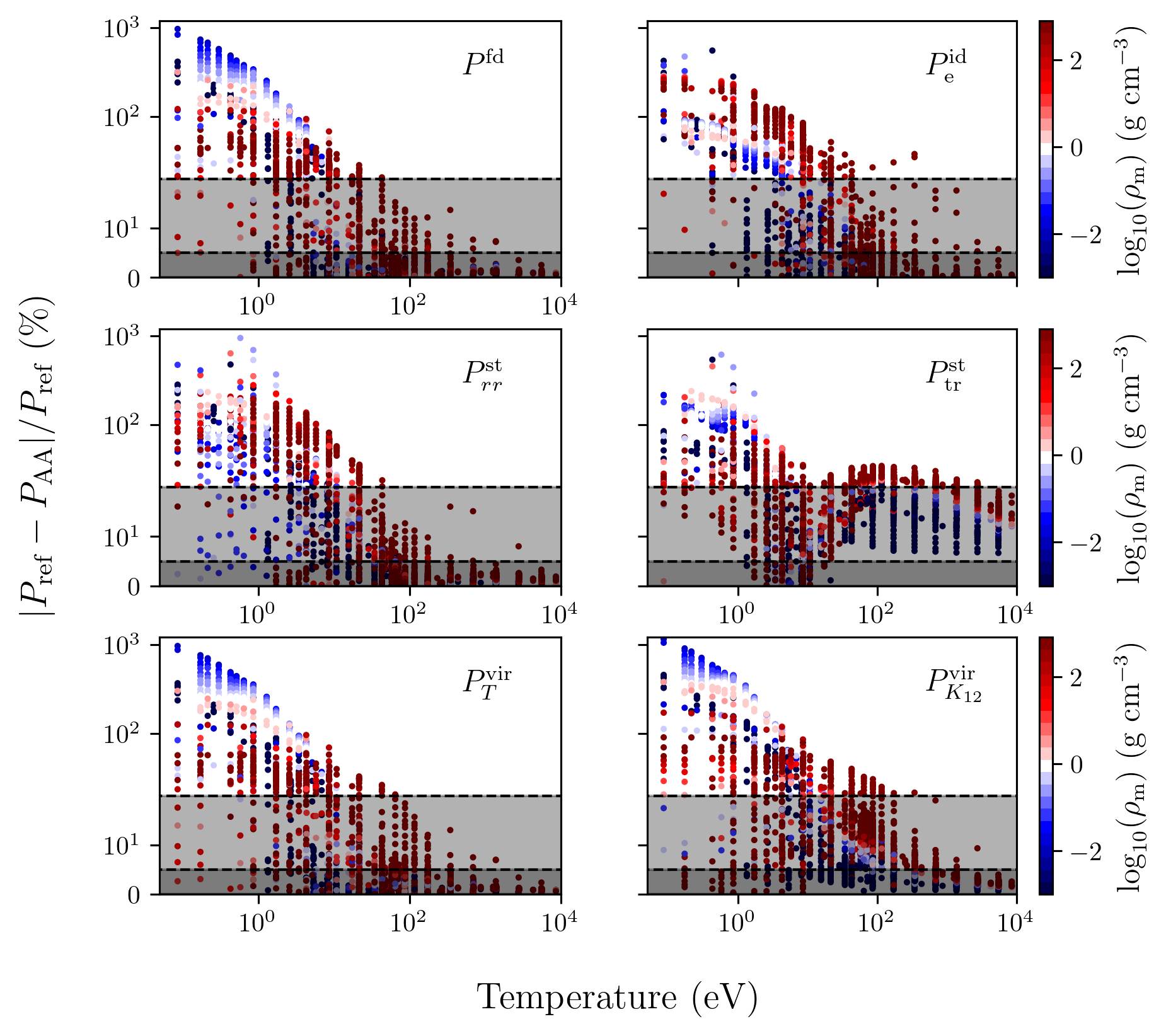}
    \caption{Absolute percentage errors (APE) in pressure between different AA methods and first-principles data \cite{Militzer_EOS_database}, as a function of temperature. A colour scale is also included to indicate the effect of mass density. Note that the $y$-axis scale is linear up to $20\%$ (with the dashed lines representing errors of $5\%$ and $20\%$ respectively), and logarithmic above $20\%$.}
    \label{fig:aa_error_temp}
\end{figure}

In Fig.~\ref{fig:aa_error_temp}, we plot the absolute percentage error (APE) \eqref{eq:MAPE} between the AA and FP pressures as a function of temperature, \change{$\epsilon_1(\Pref, P_\textrm{AA})$}, with the density also highlighted using a colour scale. Note that the $y$-axis is linear up to 20\%, and after that the scale is logarithmic. The two dashed lines (and associated shaded areas) denote errors of 5\% and 20\% respectively.

From this figure, it is clear that the AA model has large errors for low temperatures ($\lessapprox 10$ eV), independent of the approximation used for the error. The errors appear to be particularly severe when the density is also low, at least for the finite-difference and virial methods. This could partially be attributed to the asymmetry of the MAPE, where over-estimates can lead to especially large errors. However, under these conditions (low temperature and density), one would expect the electronic pressure to approach zero, with the only contribution to the pressure coming from the ions; as mass density increases for fixed temperature, we expect the electron (and ion) pressures to also increase. 

\begin{figure}
    \centering
    \includegraphics{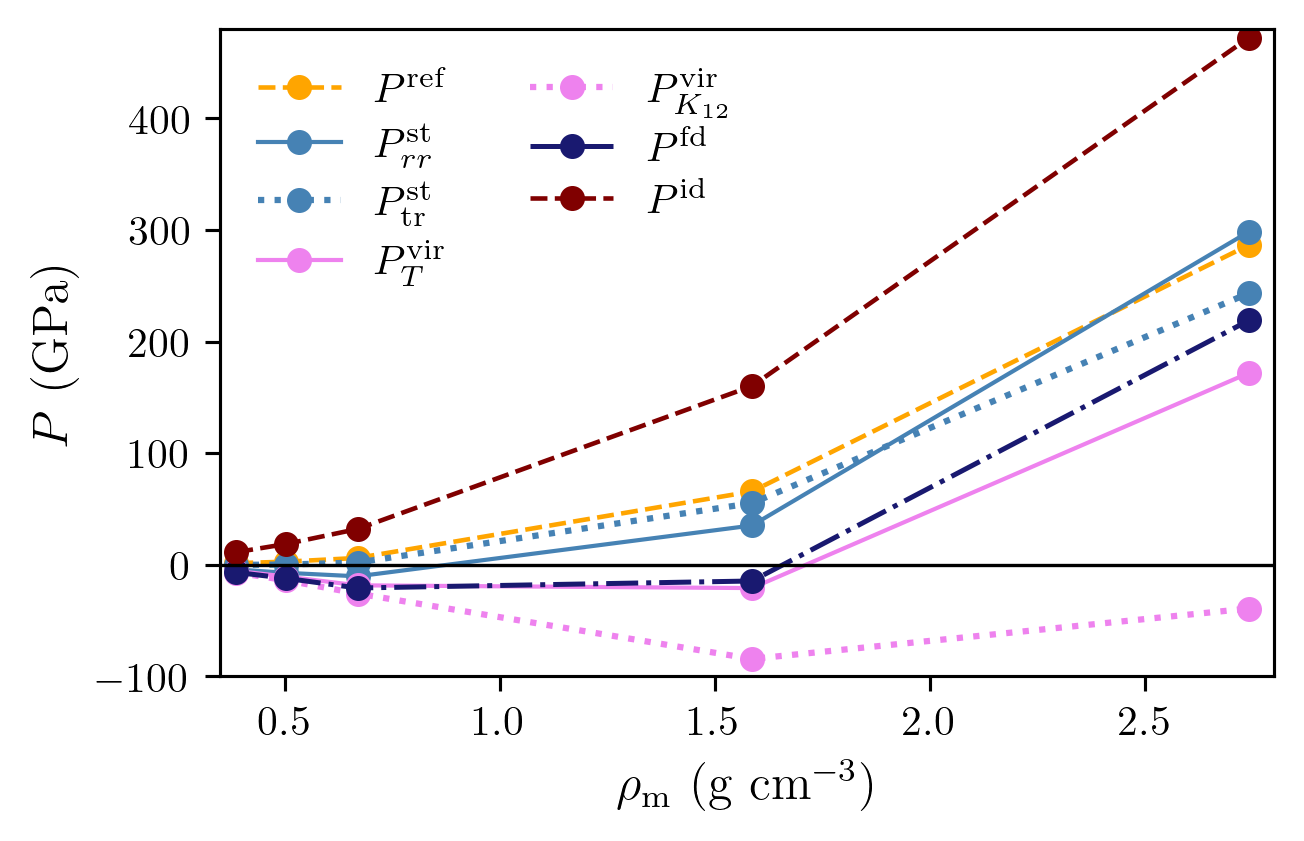}
    \caption{Pressure as a function of mass density, for Helium at temperature $T=0.043$ eV, with the various AA methods compared to the first-principles reference \cite{Militzer_EOS_database}. We observe that, for both the virial methods and the finite-difference method, the pressure actually \emph{decreases} as density \emph{increases} for the first few data points. This is both counter-intuitive, and also does not agree with the benchmark.}
    \label{fig:He_low_T_errs}
\end{figure}

However, this is not (in general) the case: consider, for example, Fig.~\ref{fig:He_low_T_errs}. In this figure, we plot the total pressure as a function of mass density for Helium, at fixed temperature $T=0.043$ eV (the lowest temperature in the FPEOS database). We see a strange behaviour in the finite-difference pressure and both virial pressures: it is negative and does not monotonically increase, as should be expected for a fixed temperature and increasing density. On the other-hand, both stress-tensor methods and the ideal pressure show the correct qualitative behaviour, with the stress-tensor results also showing good quantitative accuracy in this example. Of course, the ideal gas expression for the ions is also less accurate at low temperatures, particularly as the density increases; it is difficult to say how much the error in the total pressure is affected by errors in the ion and electron pressures, but it is likely both are incorrect to some extent.

Nevertheless, for all the pressure approximations with the exception of $\Psttr$, the AA model improves steeply with increasing temperature. This is a positive finding, because we would expect the AA model to be most accurate at higher temperatures, due to the decreasing importance of quantum effects. The trend appears to be most strong for the finite-difference and the $\Pvirt$ virial methods. A somewhat puzzling observation is the systematic errors of $\sim10-20\%$ in the high-temperature region for the $\Psttr$ method. This is in contrast to the results of Refs.~\cite{Pressure_warm_hot} and \cite{Carbon_ionization}, in which this method agreed very well with DFT-MD simulations under similar conditions. This could be related to the differences in the AA model we used compared to the model of Refs.~\cite{Pressure_warm_hot} and \cite{Carbon_ionization}.

\begin{figure}
    \centering
    \includegraphics{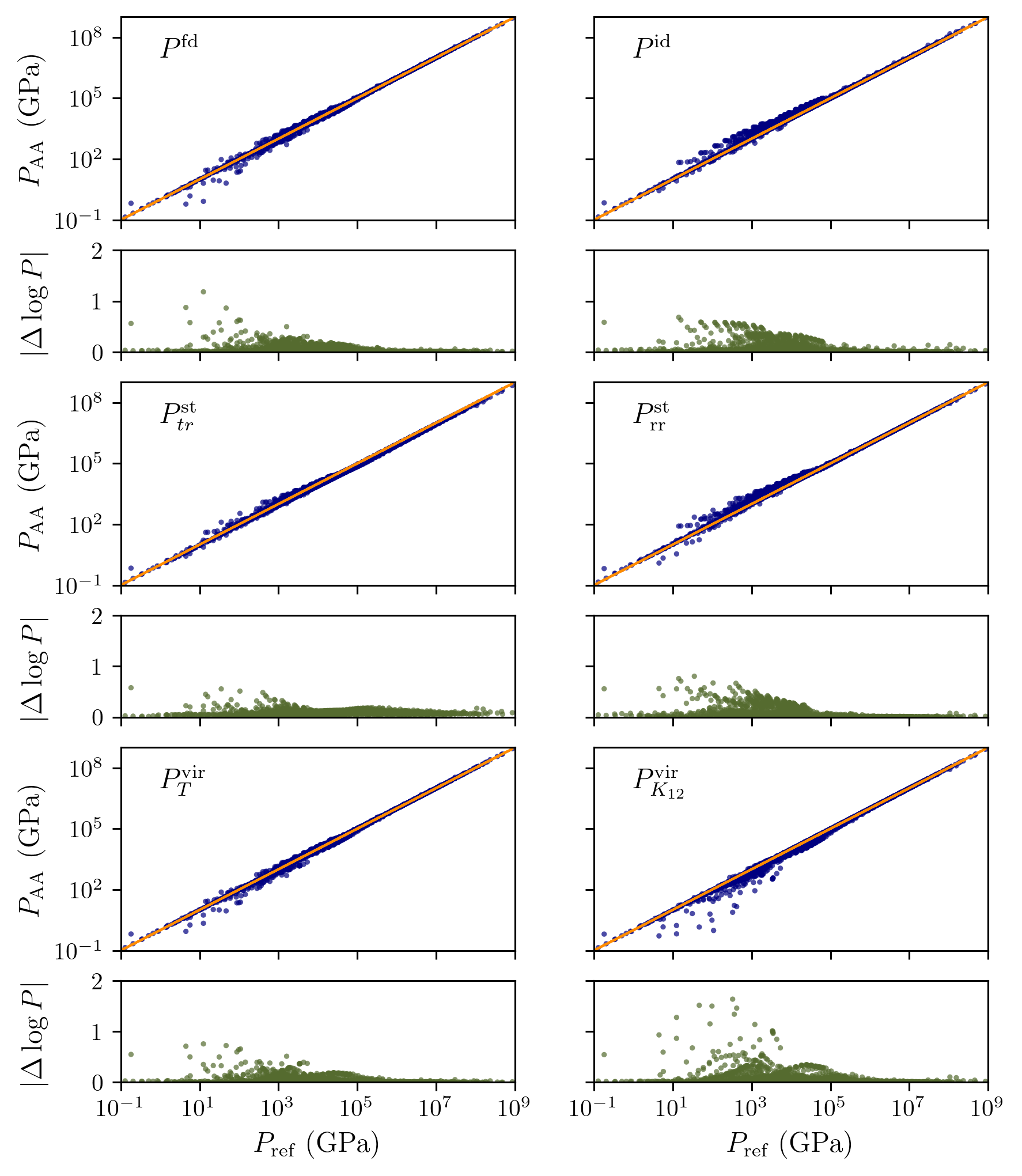}
    \caption{AA pressures (various methods) vs first-principles data \cite{Militzer_EOS_database}, plotted on a $\log-\log$ scale. The absolute log errors (ALEs) \eqref{eq:MALE} are also shown in the smaller sub-plots. In general, we observe the tendency of the AA model to perform better at higher-pressures.}  
    \label{fig:aa_error_log}
\end{figure}

Next, in Fig.~\ref{fig:aa_error_log}, we plot the AA pressures against the reference FPEOS pressure, on a logarithmic scale. In the lower panels of this plot, we plot the absolute log errors (ALE) \eqref{eq:MALE} between the AA and reference pressures, \change{$\epsilon_2(\Pref, P_\textrm{AA})$}. In this figure, it is clear that, for high pressures (high material density and/or temperatures), the AA results are in close agreement with the FPEOS reference for almost all the methods, but in particular for the virial and finite-difference methods. As was observed in Fig.~\ref{fig:aa_error_temp} for the APE, upon close observation, the stress-tensor results \change{$\Psttr$ appear to be systematically lower than the reference results at high pressures ($P\gtrapprox 10^5$)}. However, this behaviour is more noticeable when the APE is plotted (compared to the ALE); this indicates the importance of considering different error metrics when comparing pressures across a large range of temperatures and densities.

In Table~\ref{tab:AA_summary}, we collect some error metrics for all the different AA methods. From this table, we notice that the $\Pfd$, $\Pstrr$, $\Pvirt$ and $\Pid$ methods yield quite similar results for the majority of the error metrics. The $\Psttr$ and $\Pvirk$ show greater differences; in particular, they both yield a far smaller fraction of results within a $5\%$ error (by an order of magnitude in the case of $\Psttr$) and also have larger adMALE values, demonstrating bigger errors at higher pressures. This table further demonstrates the importance of considering a range of error metrics; for example, $\Psttr$ has a lower MAPE than $\Pfd$ and $\Pvirt$, but performs worse (and typically far worse) in all the other error metrics. 

\begin{table}[]
    \centering
    \begin{tabular}{ccccccc}
        \toprule
         & $\Pfd$ & $\Pstrr$ & $\Psttr$ & $\Pvirt$ & $\Pvirk$ & $\Pid$ \\ \midrule
        MAPE & 33 & 27 & 29 & 34 & 48 & 26 \\
        MALE & 8.7 & 8.7 & 11 & 8.8 & 13 & 10 \\
        adMALE & 21 & 23 & 39 & 21 & 35 & 26 \\
        $f_{20}$ & 75 & 73 & 55 & 76 & 72 & 72 \\
        $f_{5}$ & 64 & 55 & 5.4 & 60 & 24 & 55 \\
        \bottomrule
    \end{tabular}
    \caption{Aggregate error metrics for the AA methods, compared to the FPEOS dataset \cite{Militzer_EOS_database}. See Table~\ref{tab:error_metrics} for the definitions of the errors. Note that here, and in all subsequent tables, the MALE and adMALE results are multiplied by a factor of 100.}
    \label{tab:AA_summary}
\end{table}

To finish our analysis of the pressure results with the AA model, consider Table~\ref{tab:AA_summary_high_T}. This table is analagous to Table~\ref{tab:AA_summary}, however we now only consider data with temperature above $10$ eV. This quantifies the big improvements at higher temperatures seen in Fig.~\ref{fig:aa_error_temp}, and implicitly in Fig.~\ref{fig:aa_error_log}: in particular, the $\Pfd$, $\Pstrr$, $\Pvirt$, and $\Pid$ methods all have MAPEs of $\sim 4\%$, which may be considered within the bounds of experimental uncertainties. Above $10$ eV, all these methods display very similar results, and significant improvements to when all temperatures were considered. The $\Pvirk$ method also shows improvements across all metrics, but peforms somewhat worse overall than the aforementioned methods. However, as expected from Fig.~\ref{fig:aa_error_temp}, the $\Psttr$ method does not show any improvement (in fact the results are arguably worse) when temperatures below 10 eV are eliminated.

\begin{table}[]
    \centering
    \begin{tabular}{ccccccc}
        \toprule
         & $\Pfd$ & $\Pstrr$ & $\Psttr$ & $\Pvirt$ & $\Pvirk$ & $\Pid$ \\ \midrule
        MAPE & 2.6 & 4.3 & 18 & 3.1 & 8.0 & 3.9 \\
        MALE & 1.1 & 1.8 & 8.7 & 1.4 & 3.8 & 1.6 \\
        adMALE & 5.2 & 7.4 & 43 & 6.3 & 17 & 6.9 \\
        $f_{20}$ & 98 & 97 & 60 & 99 & 94 & 79 \\
        $f_{5}$ & 87 & 78 & 1.6 & 83 & 46 & 96 \\
        \bottomrule
    \end{tabular}
    \caption{Aggregate error metrics for the AA methods, compared to the FPEOS dataset\cite{Militzer_EOS_database}. The difference between this table and Table~\ref{tab:AA_summary} is that here we only consider data with temperature above 10 eV, whereas in Table~\ref{tab:AA_summary} the whole dataset was considered.}
    \label{tab:AA_summary_high_T}
\end{table}

\subsection{FPEOS database of Militzer \emph{et al.}: Neural network results}\label{sec:results_nn_fpeos}

Now, we consider results for the neural networks trained by the procedure described in Section.~\ref{sec:nn_method}. As mentioned, we trained two networks, one with features from the output of AA calculations, and one without. The results for the neural network are the aggregate of the results over the outer five test sets of the nested CV procedure.

\begin{figure}
    \centering
    \includegraphics{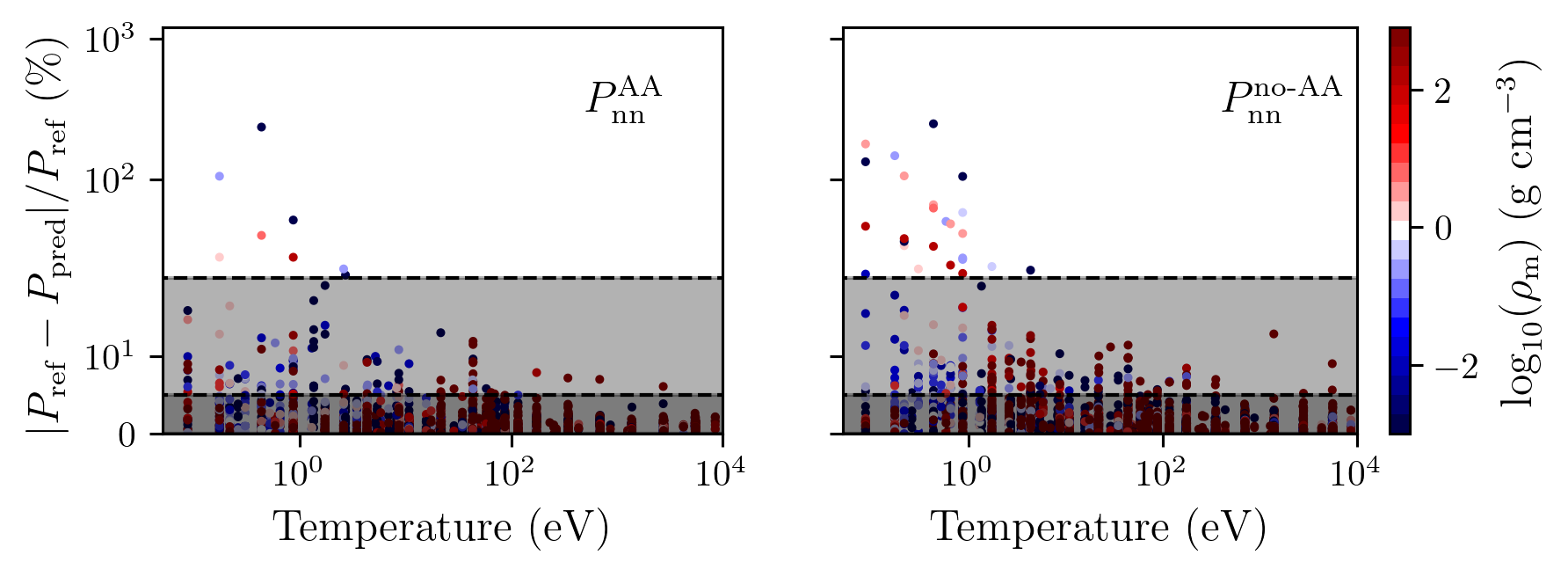}
    \caption{Absolute percentage errors (APE) for the neural network models evaluated on the FPEOS dataset \cite{Militzer_EOS_database} as a function of temperature, with a colour bar to indicate the dependence on mass density. Left: model trained with AA output as features. Right: model trained without any AA outputs.}
    \label{fig:nn_error_temp}
\end{figure}

First, consider Fig.~\ref{fig:nn_error_temp}, in which we plot the APE of the AA neural network (left panel) and the AA-free neural network (right panel). What is immediately clear from this figure, when compared to Fig.~\ref{fig:aa_error_temp}, is that both neural networks show a dramatic improvement relative to the raw AA results, at least for low-to-moderate temperatures ($T<100\ \textrm{eV}$). In particular, there are now only a handful of results with errors of over 20\%, and indeed it seems that most lie within $5\%$, even at low temperatures.

\begin{figure}
    \centering
    \includegraphics{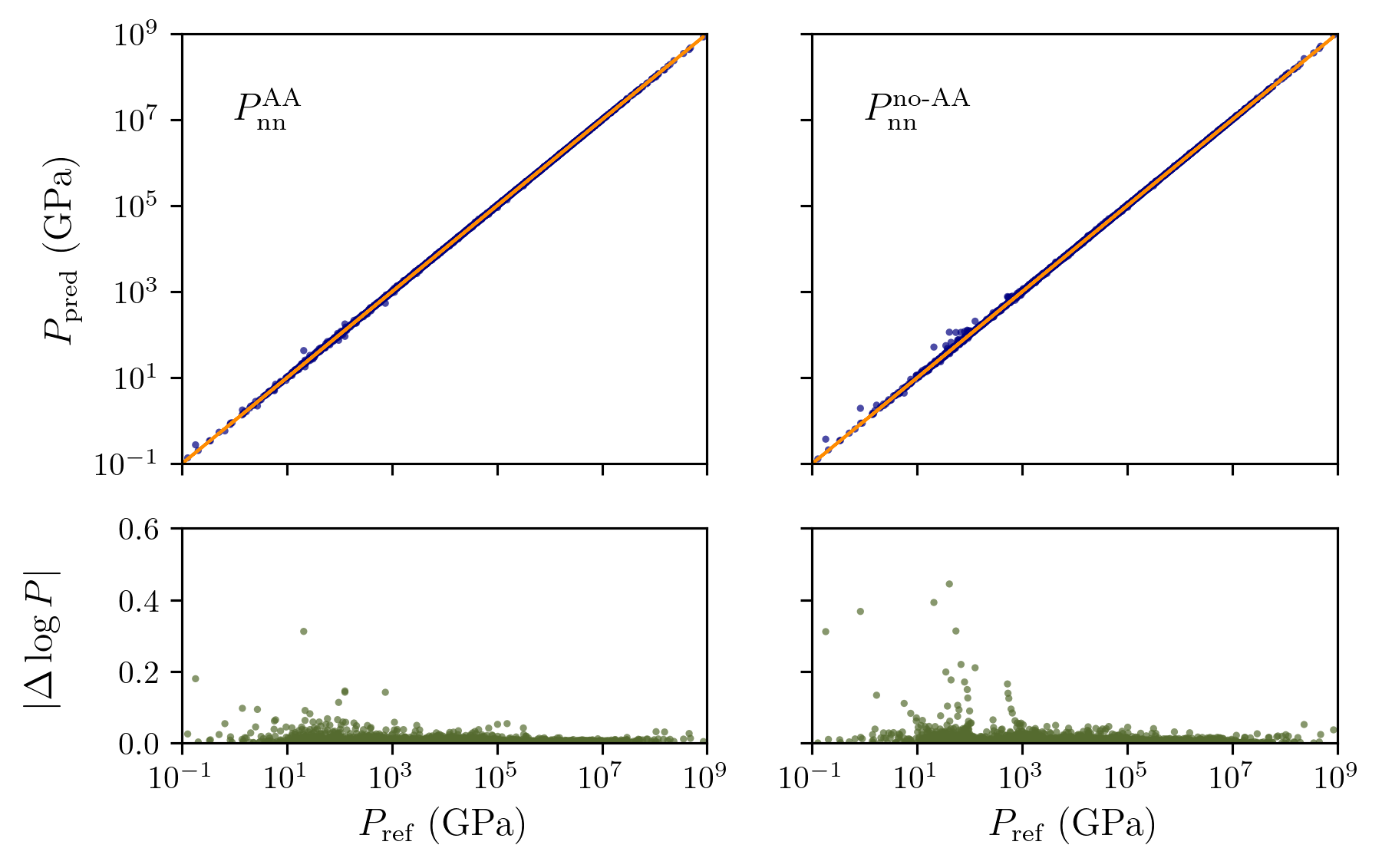}
    \caption{Pressures computed from neural network models against the benchmark pressures from Ref.~\cite{Militzer_EOS_database}. The lower panels show the logarithmic errors \eqref{eq:MALE}. Left: network trained with AA output as features. Right: network trained without any AA outputs.}
    \label{fig:nn_error_log}
\end{figure}

\begin{figure}
    \centering
    \includegraphics{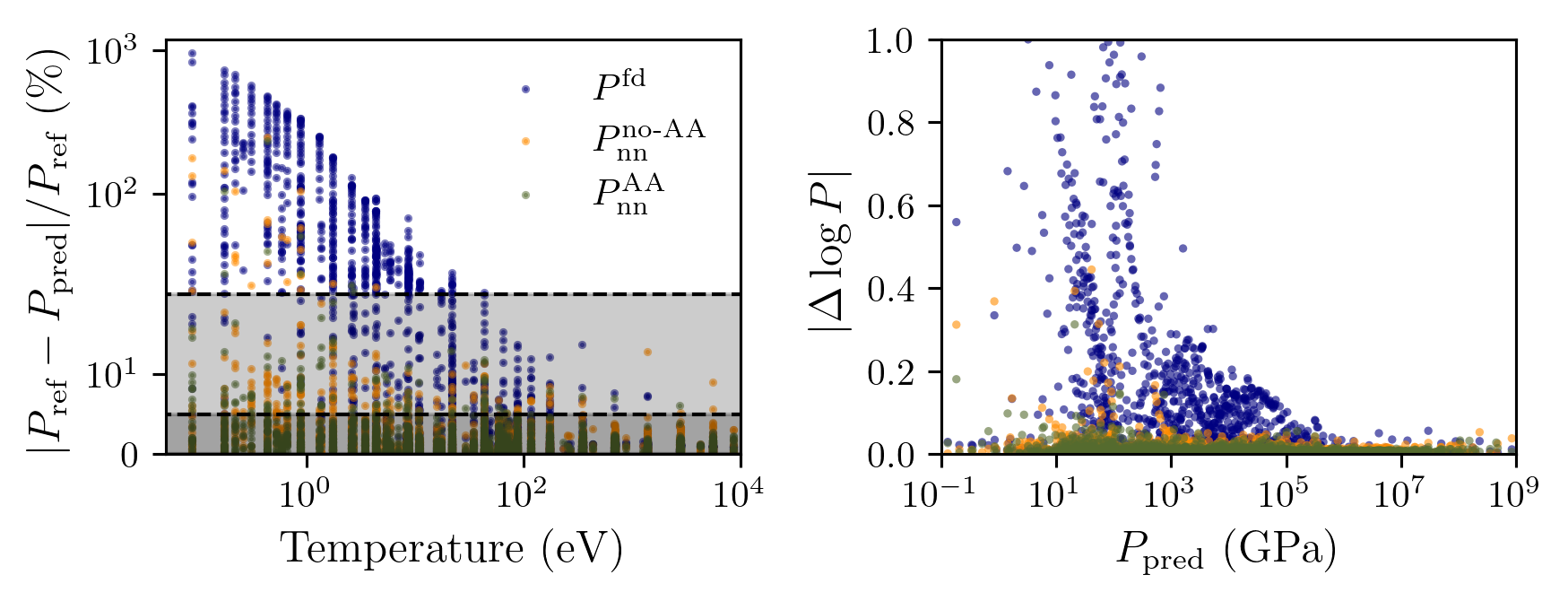}
    \caption{Comparison of errors between a raw AA model ($\Pfd$), and neural networks trained with and without AA output, evaluated on the FPEOS dataset \cite{Militzer_EOS_database}. Left: absolute percentage errors \eqref{eq:MAPE}. Right: absolute log errors \eqref{eq:MALE}.} 
    \label{fig:aa_nn_errs_comp}
\end{figure}

Fig.~\ref{fig:nn_error_log} paints a similar picture; on a logarithmic scale, it is challenging to discern any real differences between the neural-network pressures and the reference pressure. In Fig.~\ref{fig:aa_nn_errs_comp}, we directly compare the the APE (left) as a function of temperature, and the ALE (right) as a function of the reference pressure, between the two neural network models and the $\Pfd$ raw AA result. We observe clear improvements in the neural network models relative to the raw AA results in this figure, at least for $T<100\ \textrm{eV}$ and $\Pref<10^6\ \textrm{GPa}$. 

In Table.~\ref{tab:aa_CV}, we show the aggregate error metrics for the neural network models and the $\Pfd$ AA results. These numbers further demonstrate the stark improvements with the neural network models, as the error metrics are uniformly (and significantly) lower than for the raw AA results. For example, respectively only $0.4\%$ and $1.1\%$ of results now show an error of more than $20\%$ compared to the reference. Interestingly, there is not much difference between the networks trained with and without the AA outputs, although the AA network does have slightly lower average errors for all the metrics considered.

\begin{table}[]
    \centering
    \begin{tabular}{ccccccc}
        \toprule
         & $\Paa$ & $\Pnaa$ & $\Pfd$ \\ \midrule
         MAPE & 1.9 (0.2) & 2.6 (0.5) & 33 (2) \\
         MALE & 0.77 (0.1) & 1.0 (0.2) & 8.7 (0.5) \\
         adMALE & 2.5 (0.4) & 3.1 (0.7) & 21 (0.7) \\
         $f_{20}$ & 99.6 (0.3) & 98.9 (0.5) & 75 (1) \\
         $f_5$ & 94 (2) & 90 (3) & 63 (2)\\
         \bottomrule
    \end{tabular}
    \caption{Aggregate error metrics for the AA and AA-free neural networks, with AA results ($\Pfd$) also shown for comparison. The numbers in brackets for the MAPE, MALE and adMALE are the standard deviations of these quantities across the five outer cross-validation folds.}
    \label{tab:aa_CV}
\end{table}

Of course, it would come as a great surprise if the neural networks did not show a (significant) improvement to the raw AA results, since a neural network (at least in this case) is essentially a polynomial fitting to the reference data. Interpolation methods for EOS data are already a well-established tool in the WDM and plasma-physics communities. \change{In Appendix~\ref{app:interp_comp}, we evaluate the performance of direct interpolation methods provided in Ref.~\onlinecite{Militzer_EOS_database} using a similar cross-validation approach to the one used to test our neural network models. One of the conclusions we draw from Appendix~\ref{app:interp_comp} is that our method has greater capability to extrapolate to densities and temperatures outside of the training range. Another benefit of our method is that it is more global, in that it does not fit to a specific element but instead to the whole space of elements considered. Results supporting this claim will be shown in Sec.~\ref{sec:results_nn_Be}. Our work has further novelty because, to the best of our knowledge, output from AA calculations has not previously been used to supplement interpolations of first-principles (or alternative sources of high-fidelity) data. Our work is similar in spirit to Ref.~\cite{Mentzer_EOS_neural}, but it is distinguished by the inclusion of the AA outputs, besides various other differences such as the reference data, the neural network architecture, and the training and evaluation framework.}

\subsection{FPEOS Be data of Ding and Hu: Average-atom and neural network results}\label{sec:results_nn_Be}

Next, we consider the FPEOS database from Ref.~\cite{Hu_Be_EOS}, which we henceforth denote as the FP-Be dataset. In this case, all the data comes from DFT-MD calculations, with KS-DFT-MD used for temperatures up to 250,000 K ($21.5\ \textrm{eV}$), and OF-DFT-MD used for the calculations with temperatures above that. They observed that, at the temperature transition point, the results between the two methods were consistent to within $1\%$. Beryllium is the only element in this dataset. We note that, in the results we shall present, we use only the raw data and not any extrapolated or interpolated values; and like for the FPEOS data, there were a small number of high-temperature and low-density calculations which we are currently not able to perform with atoMEC.

\begin{figure}
    \centering
    \includegraphics{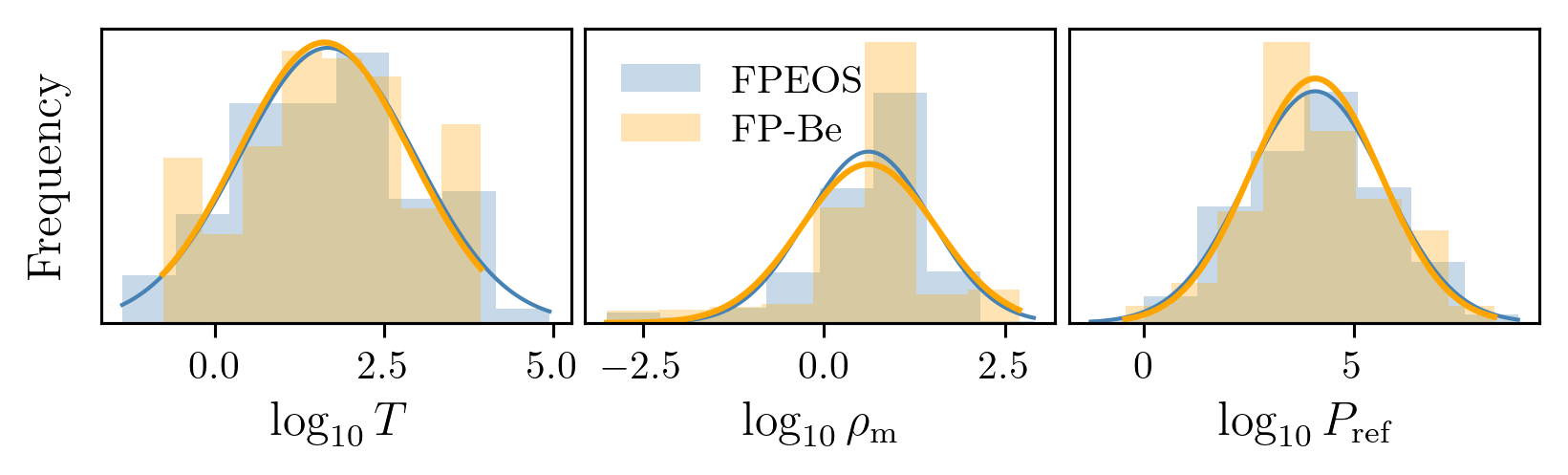}
    \caption{Distribution of temperatures (left), mass densities (middle) and reference pressures (right) in the FPEOS \cite{Militzer_EOS_database} and FP-Be \cite{Hu_Be_EOS} datasets.}
    \label{fig:Be_dist_comp}
\end{figure}

We emphasize that this dataset is completely separate from the FPEOS data used to train the neural network models, and furthermore, that Beryllium does not feature in the FPEOS dataset. Therefore this tests the ability of the neural network models to extrapolate to elements on which they were not trained, instead of the more straightforward interpolation considered in the previous section. Of course, Beryllium has an atomic number of 4, which lies inside the range $1\leq Z\leq 14$ spanned by the FPEOS dataset. We also note that the FP-Be data has a similar temperature and density distribution as the FPEOS dataset, as shown in Fig.~\ref{fig:Be_dist_comp}. In this figure, we see the temperature profiles are almost identical, and while the FP-Be data is slightly more skewed towards higher densities (and hence higher pressures), the difference is quite small.

\begin{figure}
    \centering
    \includegraphics{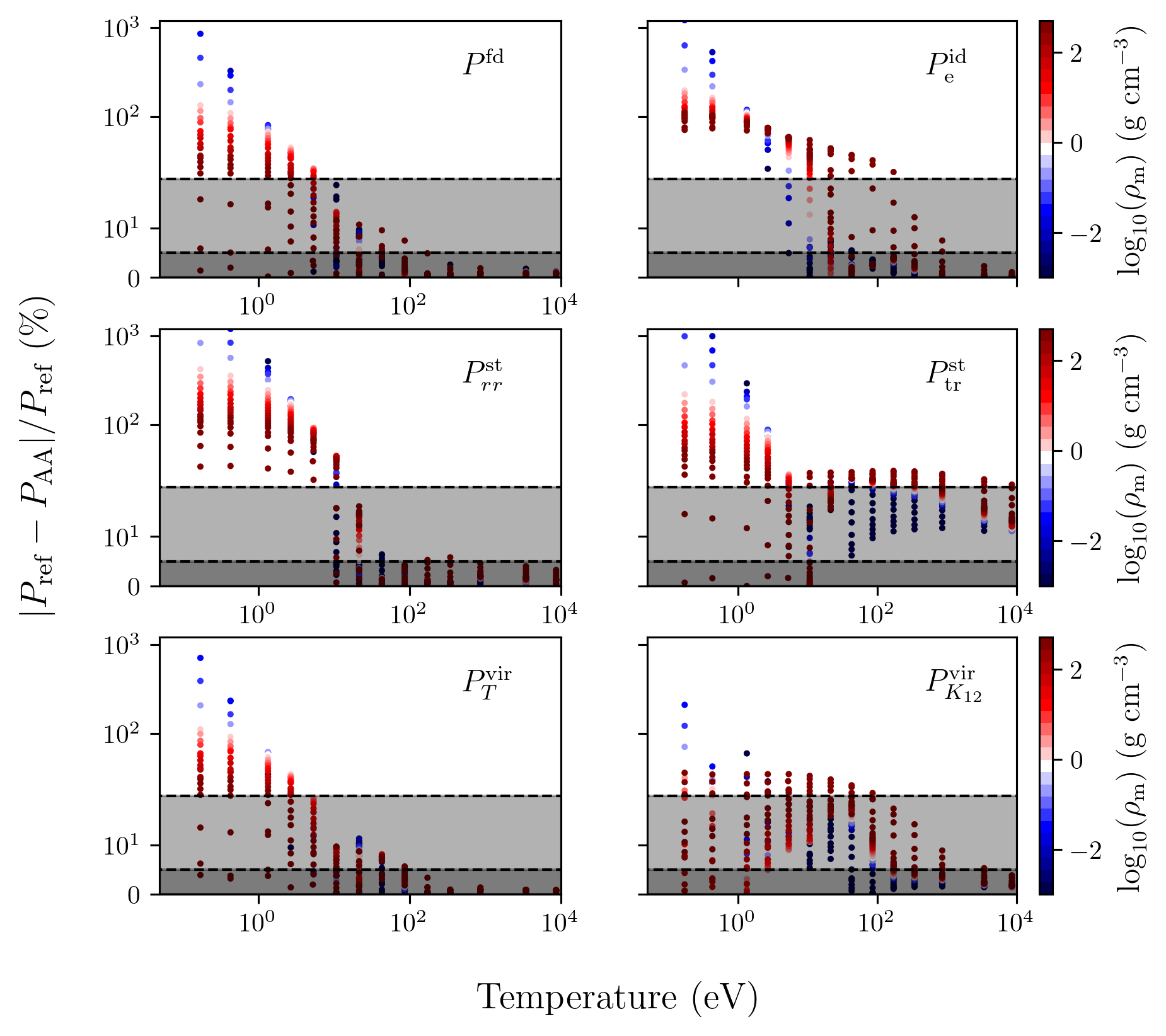}
    \caption{Absolute percentage errors (APE) in pressure between different AA methods and Be first-principles data \cite{Hu_Be_EOS}, as a function of temperature, with the colour scale indicating the dependence on mass density.}
    \label{fig:err_T_aa_Be}
\end{figure}

In Fig.~\ref{fig:err_T_aa_Be}, we compare the APEs across all the different AA models. As with Fig.~\ref{fig:aa_error_temp}, we see that all methods with the exception of $\Psttr$ tend to approach the reference pressure as temperature increases; and we again observe very large errors for $T\lessapprox 10$ eV. This observation is especially stark for the $\Pstrr$ method, in which the errors drop sharply from $\sim 100\%$ to $<5\%$ almost immediately as temperature increases beyond 10 eV.

\begin{figure}
    \centering
    \includegraphics{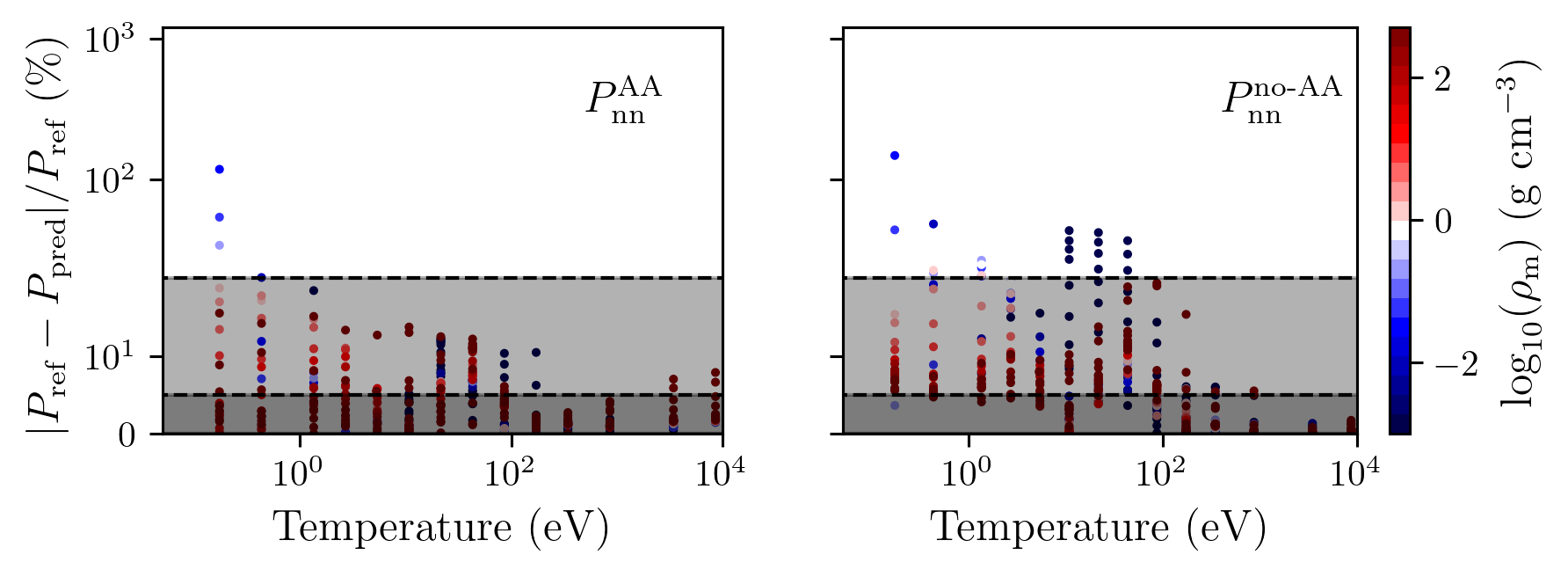}
    \caption{Absolute percentage errors (APE) for the neural network models evaluated on the FP-Be dataset \cite{Hu_Be_EOS} as a function of temperature, with a colour bar to indicate the dependence on mass density. Left: model trained with AA output as features. Right: model trained without any AA outputs.}
    \label{fig:err_T_nn_Be}
\end{figure}

Next, consider Fig.~\ref{fig:err_T_nn_Be}, in which we plot the APE for the two neural network models. \change{Both models show a clear improvement relative to the raw AA results in the low-to-mid temperature range; this is particularly true for the model trained with AA features, which has very few results with an error $>20\%$.} On the other hand, under high temperatures at which the AA models perform strongly, there is no evidence of improvement for the neural networks. However, it is worth noting that the AA results are themselves highly accurate at these temperatures, so it would be difficult for the neural networks to demonstrate much improvement.

\begin{figure}
    \centering
    \includegraphics{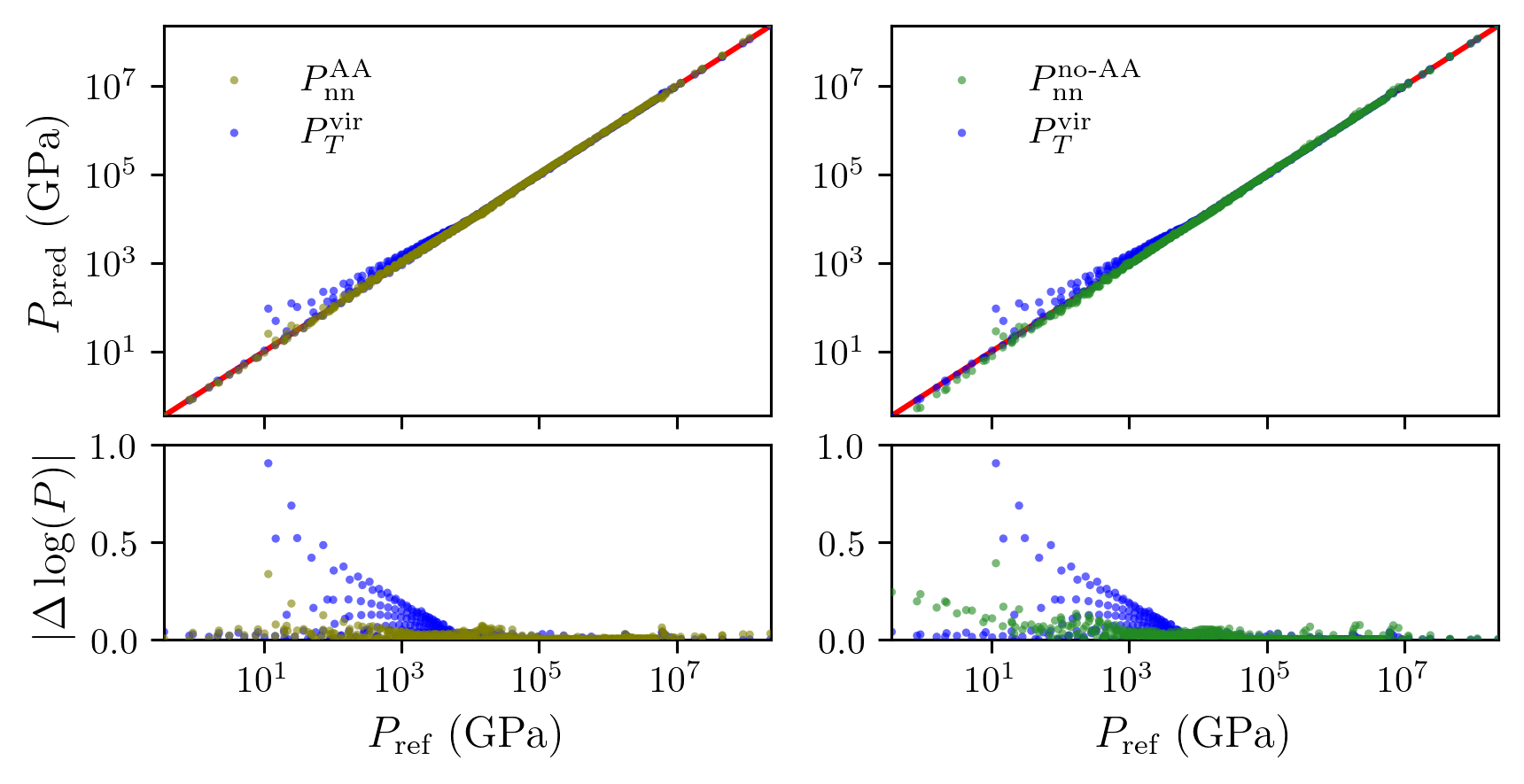}
    \caption{Pressures computed from neural network models and the $\Pvirt$ AA method, against the benchmark pressures from Ref.~\cite{Hu_Be_EOS}. The lower panels show the logarithmic errors \eqref{eq:MALE}. Left: network trained with AA output as features. Right: network trained without any AA outputs.}
    \label{fig:Be_log_nn_aa}
\end{figure}

\begin{figure}
    \centering
    \includegraphics{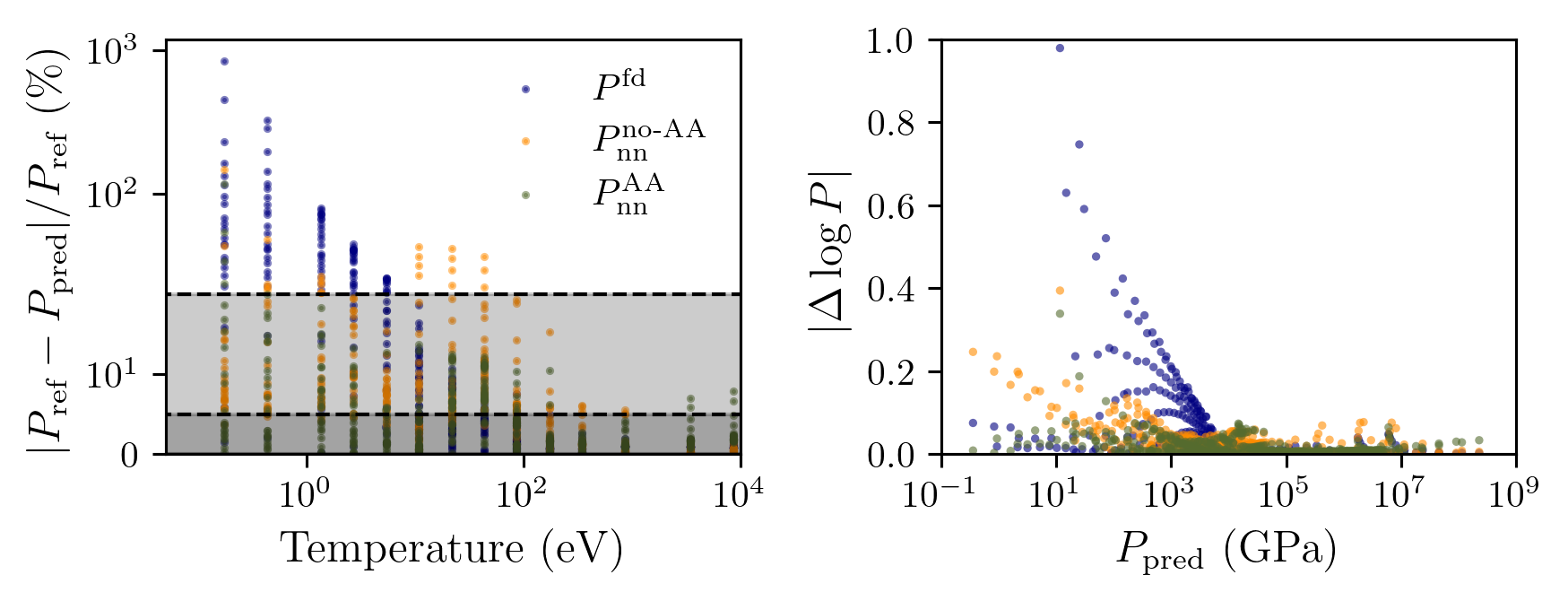}
    \caption{Comparison of errors between an AA model ($\Pvirt$), and neural networks trained with and without AA output, evaluated on the FP-Be dataset \cite{Hu_Be_EOS}. Left: absolute percentage errors \eqref{eq:MAPE}. Right: absolute log errors \eqref{eq:MALE}.}
    \label{fig:err_aa_nn_comp_Be}
\end{figure}

The above observations are further borne out when considering Figs.~\ref{fig:Be_log_nn_aa} and \ref{fig:err_aa_nn_comp_Be}. We can see, by direct comparison of the errors between the neural network and AA model (which was chosen as the virial method $\Pvirt$ in this case), that the AA result has larger errors and lower temperatures and pressures. However, it rapidly converges to the reference. Both neural networks perform relatively strongly across the full range of conditions spanned, albeit with evidence of slightly larger errors at the highest temperatures and pressures. \change{In these figures, the network trained with AA data seems markedly more accurate than the raw AA results or the network trained without AA data.}

Now consider Table~\ref{tab:AA_Be_summary}, in which we compare error metrics for the whole set of AA pressures and also the neural network predictions. Of the AA methods, $\Pfd$ and $\Pvirt$ seem to overall perform best, and also perform better than they had done on the FPEOS data. One cause could be the slightly different mass density distributions between the two datasets; the MAPE (and to a lesser extent the MALE and adMALE) can be easily influenced by large errors. The $f_5$ and $f_{20}$ scores, which are not unduly influenced by outliers, demonstrate the most consistency between the two datasets. \change{The predictions from the neural network trained with AA data show significant improvement in all error metrics relative to the raw AA results; meanwhile, the predictions from the neural network trained without AA data also show (more moderate) improvement compared to the raw AA results in all metrics aside from the $f_5$ score.}
\change{However, the improvement of the neural network models relative to the raw AA results for the FP-Be dataset is not as strong as the equivalent improvement for the FPEOS dataset. We can see this, for example, by comparing Table \ref{tab:AA_summary} and Table \ref{tab:AA_Be_summary}. There is a bigger difference in the various error metrics between the $\Pfd$ AA pressure in Table \ref{tab:AA_summary} and the neural network pressures, than there is in Table~\ref{tab:AA_Be_summary}. This is
an indication that the neural networks can interpolate to elements to which they were not
trained on, but that the performance will drop slightly as a result.}

\begin{table}[]
    \centering
    \begin{tabular}{ccccccccc}
        \toprule
         & $\Pfd$ & $\Pstrr$ & $\Psttr$ & $\Pvirt$ & $\Pvirk$ & $\Pid$ & $\Paa$ & $\Pnaa$ \\ \midrule
        MAPE & 18 & 68 & 44 & 15 & 11 & 36  & 4.5 & 7.0\\
        MALE & 5.4 & 13 & 13 & 4.8 & 5.2 & 9.9 & 1.9 & 3.1 \\
        adMALE & 15 & 37 & 46 & 13 & 19 & 31 & 6.6 & 8.9 \\
        $f_{20}$ & 79 & 67 & 46 & 83 & 88 & 65 & 98.8 & 95 \\
        $f_{5}$ & 59 & 59 & 5.5 & 57 & 38 & 55 & 70 & 47\\
        \bottomrule
    \end{tabular}
    \caption{Aggregate error metrics for the AA methods, compared to the FP-Be dataset \cite{Hu_Be_EOS}. See Table~\ref{tab:error_metrics} for definitions of the error metrics.}
    \label{tab:AA_Be_summary}
\end{table}

As a final analysis, consider Table~\ref{tab:AA_Be_summary_high_temp}, in which we compare error metrics for the AA methods against the neural networks, but this time including only temperatures above 10 eV in our analysis. We see that the AA results are very comparable to Table~\ref{tab:AA_summary_high_T}; here, the $\Pfd$ and $\Pvirt$ methods show the best performance when all metrics are considered, with $\Pid$ and $\Pstrr$ slightly behind. Again, the $\Psttr$ method is a clear outlier. Interestingly, the neural network results do not seem to show improvement relative to the best AA methods in Table~\ref{tab:AA_Be_summary_high_temp}. This suggests that, when considering elements on which training has not been performed, and temperatures above 10 eV, the base AA model (with the $\Pvirt$ or $\Pfd$ method) may be more suitable than either of the neural network models. Of course, this conclusion does not apply when considering the full temperature range, or for elements present in the training data.

\begin{table}[]
    \centering
    \begin{tabular}{ccccccccc}
        \toprule
         & $\Pfd$ & $\Pstrr$ & $\Psttr$ & $\Pvirt$ & $\Pvirk$ & $\Pid$ & $\Paa$ & $\Pnaa$ \\ \midrule
        MAPE & 2.8 & 5.2 & 18 & 2.9 & 8.7 & 6.2 & 3.4 & 5.1\\
        MALE & 1.2 & 2.1 & 8.7 & 1.3 & 4.1 & 2.4 & 1.5 & 2.4 \\
        adMALE & 4.3 & 7.7 & 4.1 & 4.7 & 18 & 11 & 6.1 & 7.1 \\
        $f_{20}$ & 100 & 93 & 57 & 100 & 95 & 89 & 100 & 96 \\
        $f_{5}$ & 80 & 82 & 6.0 & 76 & 47 & 76 & 76 & 63 \\
        \bottomrule
    \end{tabular}
    \caption{Aggregate error metrics for the AA methods, compared to the FP-Be dataset \cite{Hu_Be_EOS}. The difference between this table and Table~\ref{tab:AA_Be_summary} is that here we only consider data with temperature above 10 eV, whereas in Table~\ref{tab:AA_Be_summary}, the whole FP-Be dataset was considered. See Table.~\ref{tab:error_metrics} for definitions of error metrics.}
    \label{tab:AA_Be_summary_high_temp}
\end{table}

\section{Conclusions}

In this paper, we sought to develop methods to accurately and efficiently compute equation-of-state (EOS) data, specifically the pressure, in the warm dense matter (WDM) regime. %
\change{First-principles} methods to compute accurate EOS data under WDM conditions --- namely, density-functional theory molecular-dynamics (DFT-MD) and path integral Monte Carlo (PIMC) --- are well known. However, these methods are far too expensive to generate on-the-fly data for hydrodynamic simulations. \change{AA calculations, on the other hand, often take only a few seconds \cite{Massacrier_bands_2021}, and are fast enough to run in-line with hydrodynamic simulations \cite{Hansen_fast_AA, XSN_ref_2}.}

In \change{the first part of the} paper, we first compared the performance of an AA model to the FPEOS dataset of Militzer \emph{et al.} \cite{Militzer_EOS_database}. Within this model, we considered various methods for calculating pressure, which were detailed in Section~\ref{sec:AA_P}. We observed that, whilst all the methods have significant errors at relatively low temperatures ($\lessapprox 10$ eV), they almost unanimously converge to the reference results as temperature increases. For example, as seen in Table~\ref{tab:AA_summary_high_T}, if only temperatures above $10$ eV are included, the mean absolute percentage error (MAPE) is just 2.6\%, and 98\% of the results lie within 20\% of the reference. These errors are of the same order of magnitude as could be expected from experimental measurements.
A similar performance was observed when the AA pressures were compared with the FP-Be database of Ding and Hu \cite{Hu_Be_EOS}.

\change{However}, the primary goal of this paper was to investigate the application of neural networks to EOS calculations: to this end, we developed two neural networks, using the procedure described in Section~\ref{sec:nn_method}. The difference between the two networks was that one used output data from AA calculations --- see Table~\ref{tab:init_features} --- as input features to the network. The other just used fundamental physical data such as the temperature and mass density. Both these models demonstrated the ability to accurately interpolate the pressure, with MAPEs of 1.9\% (AA network) and 2.6\% (AA free network) across the whole temperature range respectively.

\change{There are, of course, classical interpolation schemes which can be used to rapidly compute EOS data based on the first-principles datasets used in this paper. However, our method has several advantages relative to these schemes. Most importantly, our neural network models are global (rather than being confined to a specific element). This means that, firstly, our methods can be applied to elements outside of the original training set. We tested our pre-trained networks on the FP-Be dataset \cite{Hu_Be_EOS} to evaluate their performance on an element not in the training set. The network trained with AA features demonstrated significant improvement relative to the raw AA results in the low-to-mid temperature range, and maintained a roughly equal performance with the AA model at temperatures above 10~eV, when the AA model is itself highly accurate. The second advantage of our methods being global is that they can be extrapolated beyond the boundaries of the interpolation tables. This is discussed in detail in Appendix~\ref{app:interp_comp}, in which we compare direct interpolation of the FPEOS tables with our neural network models. Furthermore, though beyond the scope of this paper, we note that neural network models hold the promise of extensibility to non-local thermal equilibrium (NLTE) modelling, as outlined in Appendix~\ref{app:interp_comp}.}

Overall, the results in this paper indicate that --- with a good choice of method for calculating the pressure --- AA models generally show accurate agreement with DFT-MD and PIMC benchmarks at moderate-to-high temperatures ($\gtrapprox 10$ eV). In fact, it is at these temperatures that AA models are most useful, because DFT-MD calculations are more computationally feasible at low temperatures. 

Moreover, we have developed neural network models which, especially if trained with output data from AA calculations, are highly effective interpolation tools for first-principles EOS data. There is even evidence that these models, in particular the model trained with AA outputs, can be applied successfully to elements on which they were not trained: for example, the networks demonstrate significant improvements relative to raw AA results at low temperatures ($<10$ eV). This offers a promising solution for a global EOS method that is both accurate and efficient throughout the WDM regime.

\section*{Data Availability}

The AA results and final trained networks can be downloaded from this link: \href{https://rodare.hzdr.de/record/2289}{rodare.hzdr.de/record/2289}. The code required to run the AA calculations, and train and test the neural networks, can be accessed here: \href{https://github.com/atomec-project/neuralEOS}{github.com/atomec-project/neuralEOS}. The FPEOS database of Militzer \emph{et al.} can be downloaded from the Supplementary Material of Ref.~\onlinecite{Militzer_EOS_database}. The FPEOS Be data from Ding and Hu can be requested from the authors of Ref.~\cite{Hu_Be_EOS}, as stated in that paper.

\section*{Acknowledgements}

We thank S. Hu for providing us with the FP-Be dataset \cite{Hu_Be_EOS}. This work
was partially supported by the Center for Advanced Systems
Understanding (CASUS), which is financed by Germany’s
Federal Ministry of Education and Research (BMBF) and by
the Saxon state government out of the State budget approved
by the Saxon State Parliament.
We declare that there are no conflicts of interest.

\appendix
\section{Comparison of neural network models with FPEOS direct interpolation}\label{app:interp_comp}

In hydrodynamic simulations, EOS data is required on-the-fly, and therefore interpolation of pre-computed EOS tables is a common practice. In the FPEOS database used in this paper \cite{Militzer_EOS_database}, classical interpolation routines are provided with the data. In this appendix, we compare the accuracy and computational load of these interpolation routines with our neural network models.

First, let us consider the computational load of the interpolation routines relative to our neural network approaches. After the neural networks have been trained --- which is a one-time procedure as described in Section~\ref{sec:nn_method} --- performing inference with the networks, i.e. applying the trained networks to new data, is effectively instantaneous, just like standard interpolation routines.

\begin{figure}
    \centering
    \includegraphics{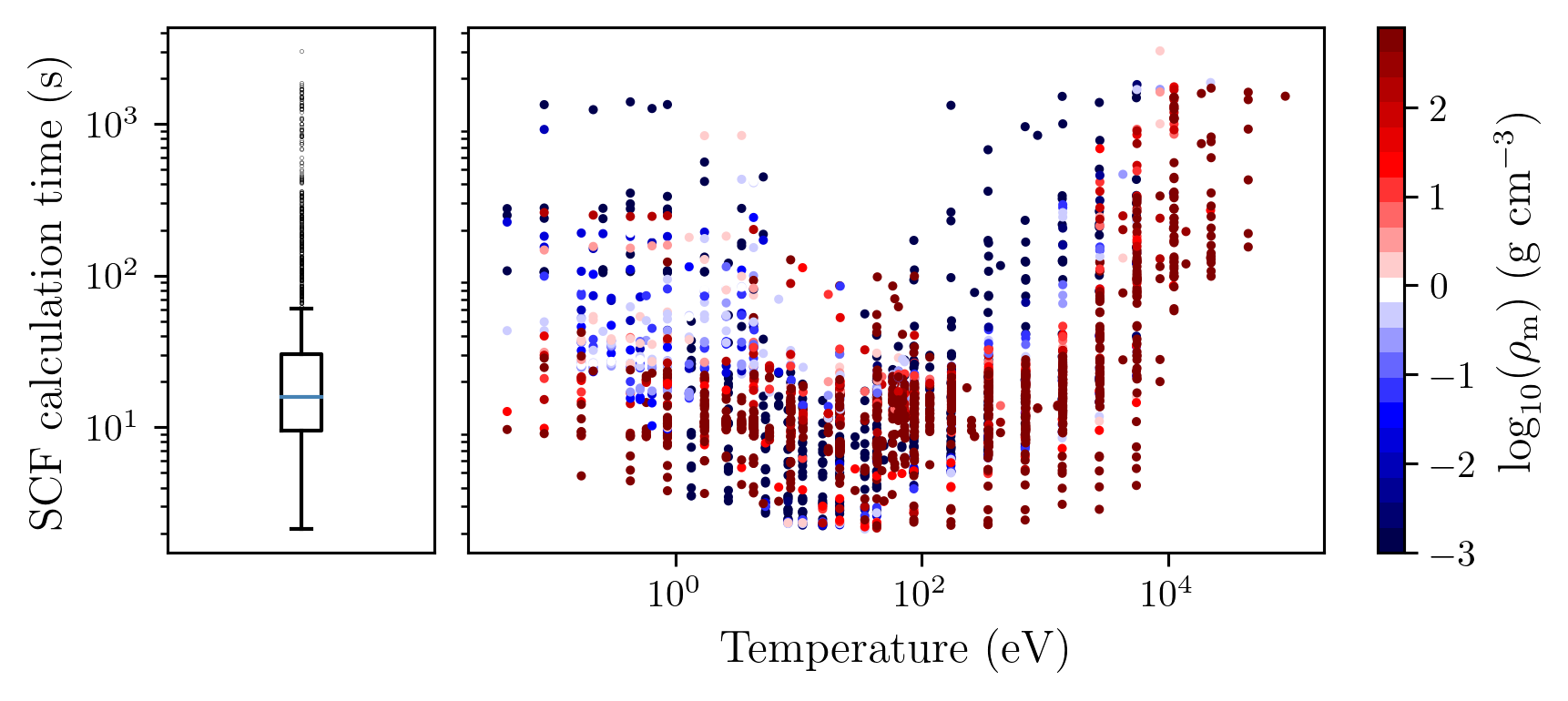}
    \caption{Average-atom SCF calculation times. The left panel is a box plot (showing the median time, interquartile ranges, limits (without outliers), and outliers. The right panel shows the time taken per calculation, as a function of temperature and density.}
    \label{fig:AA_timings}
\end{figure}

However, the network which uses AA data as input features has an extra computational load, namely the computation of the AA data. As mentioned in the conclusions of this paper, there exist AA codes which are fast enough to be run inline with hydrodynamic simulations \cite{Massacrier_bands_2021, Hansen_fast_AA, XSN_ref_2}. On the other hand, the atoMEC code we have used is not (in general) fast enough to generate on-the-fly data for hydrodynamic simulations. Fig.~\ref{fig:AA_timings} shows the AA SCF calculation times for all the calculations in the FPEOS dataset. The median time is 16 seconds, with the 25\% and 75\% quartiles being 10 and 31 seconds respectively. Some calculations, however, are much more expensive (taking over 600 seconds, i.e. 10 minutes). The cost increases when more orbitals are required, i.e. when the density is lower or the temperature higher (the cost is also dependent on convergence parameters such as the grid size). 

Firstly, we note that development of the atoMEC code only began two years ago, and so far was not focussed on performance. We imagine that the code could be sufficiently optimized so that it performs as well as the codes mentioned in Refs.~\cite{Massacrier_bands_2021, XSN_ref_2}. Secondly --- and more importantly --- it is in fact not necessary to run an AA code inline to generate the input data for the neural networks. In fact, we are developing an approach to interpolate AA data across the full thermodynamic spectrum that can be covered by the atoMEC code, which will grant instantaneous access to AA data. 

As an aside, we note that this approach has the additional advantage that it can be extended to non-local thermal equilibrium (NLTE) conditions. Solving the collisional-radiative equations requires spectral information as input data, which is most accurately obtained from a fully-quantum method such as an AA code. Under NLTE conditions, multiple configurations must be accounted for, and it is therefore not possible to run even the fastest AA codes inline; the number of configurations must be truncated, severely limiting the accuracy. Furthermore, by using a neural network to interpolate the AA data, rather than a classical interpolation scheme, the complexity of the system is reduced and we expect to be able to scale the method to NLTE.

Next, we compare the accuracy of direct interpolation of the FPEOS data with the AA neural network model. Firstly, we recall that the FPEOS data can only be interpolated on a species-by-species basis, and therefore the interpolation routines can only be used for elements present in the dataset. A major advantage of our neural network approach is that it was succesfully applied, for example, to Beryllium, which cannot be done with the FPEOS routines.

\begin{figure}
    \centering
    \includegraphics{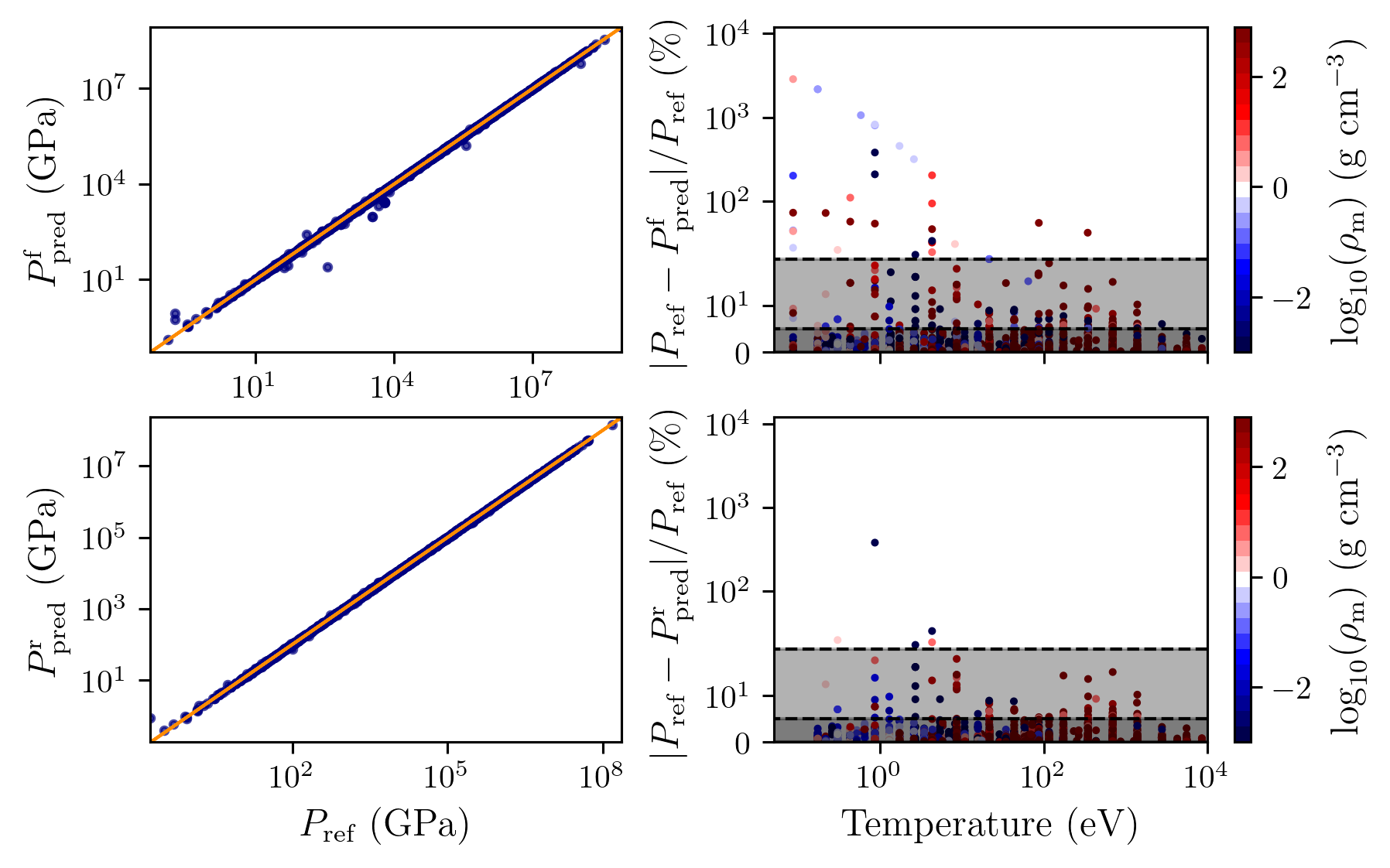}
    \caption{Comparison of predicted vs reference pressures for direct interpolation of FPEOS data. Left panels: Predicted vs reference pressures on a log-log scale. Right panels: Absolute percentage errors between predicted and reference pressures, as a function of density and temperature. Top panels: Evaluation is performed on the full test sets. Bottom panels: Evaluation is performed on the restricted test sets.}
    \label{fig:FPEOS_interp}
\end{figure}

To assess the accuracy of direct FPEOS interpolation, we used a cross-validaiton testing strategy. That is to say, we randomly split the data for each element into 5 non-overlapping ``training'' and test sets. The data was fitted (i.e. the interpolation functions constructed) using the ``training'' sets, and then the accuracy was assessed on the test sets. We note that, with this random splitting, it is possible to end up with test data that falls outside of the training range, i.e. with densities or temperatures in the test set that are fall outside of the bounds in the training set. This is no different from the neural network training, but in that case the bounds are much wider for each element, because the training is performed over all the elements present in the dataset. To fairly assess the accuracy of the FPEOS direct interpolation, we introduce additional test sets. These are restricted subsets of the original test data sets, in which data that falls outside of the training temperature and density range is removed from the test set. We test on both the full and restricted sets, and use the notation $\Ppredf$ or $P_\textrm{FPEOS}^\textrm{f}$, and $\Ppredr$ or $P_\textrm{FPEOS}^\textrm{r}$, to denote respectively when the full or restricted sets are used.

\begin{table}[]
    \centering
    \begin{tabular}{ccccccc}
        \toprule
         & $P_\textrm{FPEOS}^\textrm{r}$ & $P_\textrm{FPEOS}^\textrm{f}$ & $\Paa$ \\ \midrule
         MAPE & 1.6 (0.7) & 1300 (2000) & 1.9 (0.2) \\
         MALE &  0.64 (0.2) & 1.9 (0.7) & 0.77 (0.1)\\
         adMALE & 2.2 (0.3) & 3.5 (0.5) & 2.5 (0.4)\\
         $f_{20}$ & 99.7 (0.3) & 98 (0.3) & 99.6 (0.3) \\
         $f_5$ & 95 (2) & 93 (1) & 94 (2)\\
         \bottomrule
    \end{tabular}
    \caption{Comparison of error metrics between FPEOS direct interpolation with AA neural network model. The MAPE values for $P_\textrm{FPEOS}^\textrm{f}$ are somewhat meaningless, due to the presence of large outliers.}
    \label{tab:fpeos_vs_aa}
\end{table}

To evaluate the performance of the FPEOS networks, we first consider Fig.~\ref{fig:FPEOS_interp}. In the top panels of Fig.~\ref{fig:FPEOS_interp}, the FPEOS interpolation is evaluated on the full test sets; meanwhile, in the bottom panel, it is evaluated on the restricted test sets, which exclude data falling outside the bounds of the training data. We see that the bottom panels have far fewer large errors than the top panels; in particular, the $\Ppredf$ predictions have some extremely large errors of over $100\%$. However, in general, the vast majority of the data is closely aligned with the reference (as expected). 

Next, we consider Table~\ref{tab:fpeos_vs_aa}, in which we compare error metrics for $P_\textrm{FPEOS}^\textrm{r}$, $P_\textrm{FPEOS}^\textrm{f}$, and the AA neural network model $\Paa$. It is clear from this table that the AA model is slightly less accurate than the FPEOS interpolation on the restricted test sets, $P_\textrm{FPEOS}^\textrm{r}$. However, it is more accurate than the FPEOS interpolation on the full datasets, $P_\textrm{FPEOS}^\textrm{f}$, and in particular has fewer large outliers. Overall, we can conclude that direct interpolation of the FPEOS dataset is highly accurate; however, even when going slightly beyond the limits of the training data, the interpolation quickly becomes out of control. This is a key advantage of our method, which does not rely on an element-by-element interpolation, and therefore can be used to make predictions over a much wider range of temperatures and densities.

\bibliographystyle{unsrt}
\bibliography{main}

\end{document}